\documentclass[a4paper,11pt,pdftex]{article}
\pdfoutput=1

\usepackage{jheppub}
\usepackage[utf8x]{inputenc}
\usepackage[normalem]{ulem}
\usepackage{amssymb,amsmath}
\usepackage{bm}
\usepackage{booktabs}
\usepackage{color}
\usepackage{comment}
\usepackage{float}
\usepackage{graphicx}
\usepackage{hyperref}
\usepackage{mathrsfs}
\usepackage{url}
\usepackage{wrapfig}
\usepackage{xcolor}
\hypersetup{
colorlinks=true,
citecolor=blue,
citebordercolor=red,
linkcolor=blue,
urlcolor=blue
}

\catcode`\@=11
\def\lsim{\mathrel{\mathpalette\@versim<}}
\def\gsim{\mathrel{\mathpalette\@versim>}}
\def\@versim#1#2{\vcenter{\offinterlineskip
\ialign{$\m@th#1\hfil##\hfil$\crcr#2\crcr\sim\crcr } }}
\catcode`\@=12

\newcommand{\nn}{\nonumber\\}
\newcommand{\df}{\text{d}}
\newcommand{\p}{\partial}

\graphicspath{{./figs/}}
\newbox{\ORCIDicon}
\sbox{\ORCIDicon}{\large \includegraphics[width=0.8em]{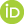}}

\makeatletter
\gdef\@fpheader{}
\makeatother

\makeatletter
\@addtoreset{equation}{section}
\makeatother

\preprint{\begin{flushright}
YGHP-22-01, KEK-TH-2418, IFT-UAM/CSIC-22-48, DESY-22-071 \end{flushright}}

\begin{document}

\title{
Coleman-Weinberg Abrikosov-Nielsen-Olesen strings
}

\author[a,b]{Minoru Eto,\,\href{https://orcid.org/0000-0002-2554-1888}{\usebox{\ORCIDicon}}}
\emailAdd{meto@sci.kj.yamagata-u.ac.jp}
\affiliation[a]{Department of Physics, Yamagata University, Kojirakawa-machi 1-4-12, Yamagata, Yamagata 990-8560, Japan}
\affiliation[b]{Research and Education Center for Natural Sciences, Keio University, 4-1-1 Hiyoshi, Yokohama, Kanagawa 223-8521, Japan}

\author[c]{Yu Hamada,\,\href{https://orcid.org/0000-0002-0227-5919}{\usebox{\ORCIDicon}}}
\emailAdd{yuhamada@post.kek.jp}
\affiliation[c]{KEK Theory Center, Tsukuba 305-0801, Japan}

\author[d,e]{Ryusuke Jinno,\,\href{https://orcid.org/0000-0003-0949-6623}{\usebox{\ORCIDicon}}}
\emailAdd{ryusuke.jinno@csic.es}
\affiliation[d]{Instituto de F\'{\i}sica Te\'orica UAM/CSIC, C/ Nicol\'as Cabrera 13-15, Campus de Cantoblanco, 28049, Madrid, Spain}
\affiliation[e]{Deutsches Elektronen-Synchrotron DESY, Notkestr. 85, 22607 Hamburg, Germany 
}

\author[b,f]{Muneto Nitta,\,\href{https://orcid.org/0000-0002-3851-9305	}{\usebox{\ORCIDicon}}}
\emailAdd{nitta@phys-h.keio.ac.jp}
\affiliation[f]{Department of Physics, Keio University, 4-1-1 Hiyoshi, Kanagawa 223-8521, Japan}

\author[g]{and Masatoshi Yamada\,\href{https://orcid.org/0000-0002-1013-8631}{\usebox{\ORCIDicon}}}
\emailAdd{m.yamada@thphys.uni-heidelberg.de}
\affiliation[g]{Institut f\"ur Theoretische Physik, Universit\"at Heidelberg, Philosophenweg 16, 69120 Heidelberg, Germany}

\abstract{
We study properties of Abrikosov-Nielsen-Olesen (ANO) strings with the Coleman-Weinberg (CW) potential, which we call CW-ANO strings.
While the scale-invariant scalar potential has a topologically trivial vacuum admitting no strings at the classical level, quantum correction allows topologically nontrivial vacua and stable string solutions. 
We find that the system of the CW potential exhibits significant difference from that of the conventional Abelian-Higgs model with the quadratic-quartic potential.
While a single-winding string is qualitatively similar in both systems, and the static intervortex force between two strings at large distance is attractive/repulsive in the type-I/II regime for both, that between two CW-ANO strings exhibits a nontrivial structure.
It develops an energy barrier between them at intermediate distance, implying that the string with winding number $n > 1$ can constitute a metastable bound state even in the type-II regime.
We name such a superconductor type-$\overline{1.5}$.
We also discuss implications to high-energy physics and cosmology.
}

\maketitle

\section{Introduction}
\label{sec:introduction}

Vortices or cosmic strings are string-like 
topological objects in quantum field theory~\cite{ABRIKOSOV1957199,Nielsen:1973cs,Rajaraman:1987,Manton:2004tk} 
playing important roles, from purely theoretical aspects such as supersymmetry~\cite{Tong:2005un,Eto:2006pg,Shifman:2007ce,Shifman:2009zz,Tong:2008qd} 
to various applications in cosmology~\cite{Kibble:1976sj,Kibble:1980mv,Vilenkin:1984ib,Hindmarsh:1994re,Vachaspati:2015cma,Vilenkin:2000jqa} 
and condensed matter systems~\cite{Mermin:1979zz,Volovik:2003fe,Svistunov:2015,Pismen,Bunkov:2000,Blatter:1994zz,Giamarchi:2002,Kawaguchi:2012ii}. 
One of the typical examples is given by 
quantum vortices or magnetic flux tubes in superconductors~\cite{ABRIKOSOV1957199,Blatter:1994zz,Giamarchi:2002}.
While conventional metallic superconductors can be well-described by the Bardeen-Cooper-Schrieffer theory, 
they can be more efficiently described 
around the critical temperature 
by the Ginsburg-Landau effective low-energy theory,
that is the Abelian-Higgs model consisting of a complex scalar field $\Phi$ representing 
a Cooper pair of electrons and a $U(1)$ gauge field $A_\mu$ representing 
magnetic fields penetrating into superconductors. 
Inside superconductors, the scalar field has a finite expectation value which yields, besides a finite scalar boson mass $m_\Phi$, a finite gauge boson mass $m_A$ as a consequence of the spontaneous breaking of the $U(1)$ gauge symmetry.  
Here, the Compton wavelengths $m_\Phi^{-1}$ and $m_A^{-1}$ give the coherence length and penetration depth, respectively.   
Hence, the scalar field represents an order parameter of the $U(1)$ gauge symmetry and the massive gauge field becomes the origin of the Meissner effect.

In the sense of the Ginzburg-Landau theory, the potential of the scalar field, $V(\Phi)$ is given as a polynomial form in terms of the order parameter, namely in the Abelian-Higgs model, a polynomial of the $U(1)$ invariant $\Phi^\dagger\Phi=|\Phi|^2$. The simplest form is given as $V_{\rm AH}(\Phi)=m^2|\Phi|^2 + \lambda|\Phi|^4$. For $m^2<0$ and $\lambda>0$, the potential has a stable vacuum $v_\Phi =\langle \Phi \rangle = \sqrt{-m^2/\lambda}$ which becomes the scale of the masses. 
Within such a setup, it has been shown that the Abelian-Higgs model contains a non-trivial static solution to equations of motion for $\Phi$ and $A_\mu$, the so-called Abrikosov-Nielsen-Olesen (ANO) vortex solution~\cite{ABRIKOSOV1957199,Nielsen:1973cs}. A key quantity characterizing the features of the ANO vortex is the ratio between the gauge and scalar masses and is denoted here by $\beta=m_\Phi^2/m_A^2$. In case of $\beta<1$, i.e. the lighter scalar boson than the gauge one, superconductors belong to the ``type-I''. In this case, vortices attractively interact and then magnetic fluxes gather. This fact implies that the type-I superconductors tend to be abruptly destroyed upon applying stronger magnetic field than a certain critical value. On the other hand,  for $\beta>1$, the gauge boson mass is lighter than the scalar one, for which superconductors belong to the ``type-II''. In this case, magnetic fluxes penetrating into 
superconductors are discretely localized in the form of the Abrikosov lattice, and thus type-II superconductors are robust against magnetic fields.
Thus, the parameter $\beta$ can be regarded as an indicator of attractive ($\beta<1$) or repulsive ($\beta>1$) interaction between ANO strings, thereby characterizing robustness of superconductors against applied magnetic field.
There have been many studies investigating the interaction between ANO strings, e.g., \cite{PhysRevB.19.4486,PhysRevB.34.6514,Speight:1996px,Bettencourt:1994kf,PhysRevB.65.224504,PhysRevB.77.144506,PhysRevB.83.054516} (see also Refs~\cite{MacKenzie:2003jp,Auzzi:2007wj,Babaev:2004hk} for vortices in extended models).

An ANO vortex string in the Abelian-Higgs model is a stable object in the sense of topological invariant which is characterized by the winding number $n$, or more precisely the first homotopy group of the vacuum. However, multivortex strings can be unstable. Their stability relies on the value $\beta$: For $\beta<1$ those are always stable, while for $\beta>1$ vortex strings with $n>1$ are unstable and decay into $n$ vortex strings~\cite{Goodband:1995rt}. In particular, for the critical coupling $\beta=1$, vortex strings 
feel neither attractive nor repulsive forces, i.e. do not interact with each other and thus multiple vortex configurations are marginally stable for arbitrary $n$. In such a case, the system is in the so-called Bogomol'nyi-Prasad-Sommerfield (BPS) state~\cite{Bogomolny:1975de,Prasad:1975kr}
corresponding to the lowest bound of the energy, allowing $2n$ moduli parameters constituting the moduli space~\cite{Tong:2005un,Eto:2006pg,Shifman:2007ce,Shifman:2009zz,Tong:2008qd}. 

In previous studies, ANO vortex solutions in the Abelian-Higgs model have been investigated intensively for the quadratic-quartic potential $V_{\rm AH}(\Phi)$~\cite{ABRIKOSOV1957199,Nielsen:1973cs,Rajaraman:1987,Manton:2004tk,Kibble:1976sj,Kibble:1980mv,Vilenkin:1984ib,Hindmarsh:1994re,Vachaspati:2015cma,Vilenkin:2000jqa}.
The above statements about the stability and forces between vortices have been established solely in the case of $V_{\rm AH}(\Phi)$ at the classical level. However, in general, the occurrence of spontaneous symmetry breaking does not restrict the potential only to the quadratic-quartic form in low-energy effective theories.
A possible example would be the Coleman-Weinberg (CW) potential~\cite{Coleman:1973jx} which induces quantum-mechanically nontrivial vacua: Starting from a scale-invariant potential $V(\Phi)=\lambda|\Phi|^4$ admitting only the trivial vacuum $\langle \Phi\rangle=0$ at the classical level, quantum corrections deform the original $|\Phi|^4$ potential logarithmically into $V(\Phi)\sim \lambda |\Phi|^4 \log (|\Phi|^2/v_\Phi^2)$ in which nontrivial vacua emerge.
The underlying mechanism is called the dimensional transmutation~\cite{Coleman:1973jx} or scalegenesis~\cite{Kubo:2015cna} in the sense that one of dimensionless couplings turns to the dimensionful parameter, i.e. the vacuum expectation value.

Coleman-Weinberg potentials have recently been attracting renewed interest in elementary particle physics, based on the argument to extend the Standard Model (SM) from the viewpoint of the classical scale invariance~\cite{Wetterich:1983bi,Bardeen:1995kv}.
Its central idea is to generate the origin of the electroweak scale via the dimensional transmutation in a scalar sector while preventing the gauge hierarchy problem (or naturalness problem).
The emergence of classical scale symmetry in the matter sector may be associated to UV theories beyond the Planck scale based on e.g. asymptotic safety~\cite{Wetterich:2016uxm} and Multi-critical Point Principle~\cite{Froggatt:1995rt,Froggatt:2001pa,Nielsen:2012pu,Haruna:2019zeu,Kawai:2021lam,Hamada:2022soj}.
A simple way to implement such an extension of the SM is introducing a classically scale-invariant Abelian-Higgs model as a ``hidden sector'' coupled to the SM via the Higgs portal coupling~\cite{Iso:2009ss,Iso:2012jn,Hashimoto:2013hta,Chun:2013soa,Kim:2019ogz,Hamada:2020vnf}, in which the radiative breaking of the hidden $U(1)$ symmetry triggers the electroweak symmetry breaking.
The Universe with such a sector may have a quite different thermal history from the Universe without it.
Indeed, such models not only allow for the formation of cosmic strings after spontaneous symmetry breaking (see e.g. Refs.~\cite{LISACosmologyWorkingGroup:2022jok,Caldwell:2022qsj} and references therein), but can also involve extremely strong first-order phase transitions~\cite{Jinno:2016knw,Kubo:2016kpb,Tsumura:2017knk,Iso:2017uuu,Chiang:2017zbz,Brdar:2018num,Marzo:2018nov,Bian:2019szo,Ellis:2020nnr}, making them a good target for ongoing and future gravitational wave searches~\cite{NANOGrav:2020qll,Desvignes:2016yex,Kerr:2020qdo,Yagi:2011wg,LISA:2017pwj,Taiji,TianQin:2020hid,Punturo:2010zz,Sesana:2019vho,AEDGE:2019nxb}.

In this paper, we investigate the basic properties of ANO strings described by the CW type potential. 
Vortices in this theory are classically unstable and quantum mechanically stable: While the classical scale-invariant potential $V(\Phi)=\lambda|\Phi|^4$ admits the trivial vacuum uniquely without any vortices, the quantum mechanically corrected CW potential admits topologically nontrivial vacua and stable vortices, which we call CW-ANO strings.\footnote{Only asymptotic behaviors of a single CW-ANO string were studied before~\cite{Morris:1992qg}.}
One of the main aims of this work is to study how the interaction between such vortex strings changes for different values of $\beta$.
In particular, we study whether there is a clear boundary between the type-I (attractive) and type-II (repulsive) regimes, and whether there is a critical coupling accompanied with the BPS state.
To this end, we consider a system of two strings and estimate its total energy as a function of the interstring distance $d$.
For the conventional Abelian-Higgs model with the quadratic-quartic potential, $\beta<1$ ($\beta>1$) corresponds to the type-I (type-II) regime for any value of $d$, and the BPS state is observed at $\beta=1$.
In contrast, we find that the CW-type strings develop an energy barrier as a function of $d$, implying that the force between them is attractive (repulsive) at small (large) distances.
Though this resembles type-1.5 superconductivity \cite{Babaev:2004hk,Moshchalkov:2009}, the attractive-repulsive relation is opposite in the CW case, and thus we call this property type-$\overline{1.5}$.
In addition, we find that the strings with multiple winding numbers can be stable or metastable, depending on the value of $\beta$.
This stability/metastability transition occurs at a critical value $\beta_c$ that is different from unity, thus making the CW-ANO string in clear contrast to the ordinary ANO string with the quadratic-quartic potential.

The organization of the paper is as follows: In Sec.~\ref{sec:model_setup}, we summarize our setups to analyze motion of ANO strings. We first give both the standard quadratic-quartic potential and the Coleman-Weinberg type potential in order to highlight differences of their structure. In Sec.~\ref{sec:axisymmetric}, the motion of a single string and the composition of the energy a string stores are investigated for both the quadratic-quartic potential and the CW-type potential by solving the equations of motion. In Sec.~\ref{sec:two_string}, we set up two-string systems and present their dynamics by solving the equations of motion for two strings numerically. In particular, there we highlight differences between the quadratic-quartic and CW cases. Sec.~\ref{sec:dc} is devoted to summarizing results and making our conclusion. In Appendix~\ref{app:universality}, we introduce other 
examples of potentials and show the string tension as a function of the distance in the two-string system.  In Appendix~\ref{app:ansatz}, we argue the validity of the superposed one-string ansatz for describing the two-strings system. Appendix~\ref{app:CW_detail} summarizes the string tensions for various values of $\beta$.

\section{Model and Setup}
\label{sec:model_setup}

In this section, we present the setup of the Abelian-Higgs model with the CW-type potential.
For comparison, the conventional quadratic-quadratic potential is also described in parallel.
After introducing the action, we describe a rescaling of quantities that greatly simplifies the following calculations.
We also comment on the justification for using the CW potential in the study of strings.

\subsection{The model}
The starting action is given by
\begin{align}
S
&=
\int \df^4 x
\left( |D_\mu \Phi|^2 - \frac{1}{4} F_{\mu \nu} F^{\mu \nu} - V(\Phi) \right),
\label{eq: starting action}
\end{align}
where $\Phi$ is a complex scalar field, $F_{\mu\nu}$ is the field strength of the $U(1)$ gauge field $A_\mu$ and $D_\mu = \partial_\mu - i g A_\mu$ is the covariant derivative. Lorentz indices are lowered or raised by the metric $\eta_{\mu\nu}={\rm diag}(1,-1,-1,-1)$.
In most of the earlier studies of the ANO string, the potential is assumed to be a simple quadratic-quartic potential (hereafter denoted by ``AH'', standing for ``Abelian-Higgs''), namely
\begin{align}
V(\Phi)&=V_{\rm AH} \equiv \lambda_{\rm AH} \Big( |\Phi|^2 - v_\Phi^2 \Big)^2,
\label{eq: SM potential}
\end{align}
with $\lambda_{\rm AH}$ the quartic coupling and $v_\Phi$ the vacuum expectation value of $\Phi$. 
In this system, the masses of the gauge and scalar bosons are given by
\begin{align}
m_\Phi^2 = 4\lambda_{\rm AH} v_\Phi^2,\qquad\qquad
m_A^2 = 2g^2v_\Phi^2,
\end{align}
respectively.

The primary goal of this work is to investigate the ANO string solutions obtained from the CW type potential whose form reads
\begin{align}
V(\Phi)&=V_{\rm CW} \equiv \lambda_{\rm eff} (|\Phi|) |\Phi|^4 - \lambda_{\rm eff} (v_\Phi) v_\Phi^4.
\label{eq: CW potential}
\end{align}
Here $\lambda_{\rm eff} (|\Phi|)$ is the effective coupling as a function of $|\Phi|$ associated with the one-loop corrections to the quartic coupling from the fields coupled to $\Phi$.
More specifically, we give
\begin{align}
\lambda_{\rm eff} (|\Phi|)
&=
\lambda_{\rm CW} \left(\ln \frac{|\Phi|^2}{v_\Phi^2} - \frac{1}{2} \right).
\label{eq:lambda_eff}
\end{align}
In general, when arbitrary numbers of (real) scalar bosons, gauge bosons and femrions are coupled to the scalar $\Phi$, $\lambda_\mathrm{CW}$ is given by
\begin{equation}
    \lambda_\mathrm{CW} = \frac{1}{16\pi^2} \left( \sum_{i=1}^{n_b} 3\kappa_i^2  + \sum_{i=1}^{n_g} g_i^4 - \sum_{i=1}^{n_f} y_i^4 \right),
\end{equation}
at the one-loop level, 
where $\kappa_i$, $g_i$, and $y_i$ are scalar-portal couplings, gauge couplings, and Yukawa couplings, respectively, while $n_b$, $n_g$ and $n_f$ are the numbers of degrees of freedom of the relevant fields.
In this work, we do not specify the fields that contribute to the logarithmic running but rather regard $\lambda_{\rm CW}$ as a free parameter.
With this parametrization \eqref{eq:lambda_eff}, the potential \eqref{eq: CW potential} has a finite global minimum located at $|\Phi| = v_\Phi$.
The masses of the gauge and scalar bosons at the minimum take the same form as the AH case
\begin{align}
m_\Phi^2 = 4 \lambda_{\rm CW} v_\Phi^2,
\qquad\qquad
m_A^2 = 2 g^2 v_\Phi^2.
\end{align}

Note that the emergence of the finite dimensionful parameter $v_\Phi$ from a scale invariant theory is the consequence of the dimensional transmutation (or scalegenesis) in the CW type potential: The existence of a finite value of $v_\Phi$ enforces a relation among dimensionless couplings (free parameters) contributing to $\lambda_{\rm CW}$, transmuting one of the free dimensionless parameters into a dimensionful parameter.

\begin{figure}
\begin{center}
\includegraphics[width=0.48\columnwidth]{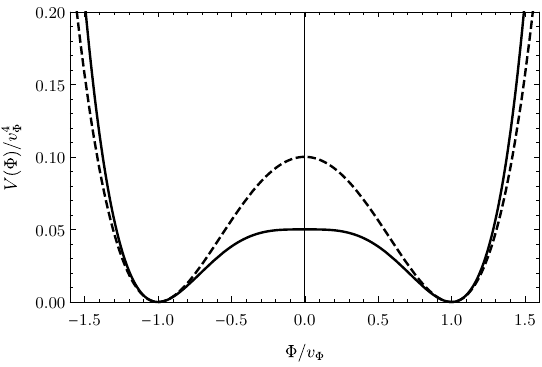}
\caption{\small
Comparison of potentials.
The solid and dashed lines indicate $V_{\rm CW}$ and $V_{\rm AH}$, respectively.
The potentials are normalized to zero at $\Phi = v_\Phi$.
We set $\lambda_{\rm AH} = \lambda_{\rm CW} = 0.1$ in this plot.
}
\label{fig: comparison of potentials}
\end{center}
\end{figure}

We here stress an important difference between Eq.~\eqref{eq: SM potential} and Eq.~\eqref{eq: CW potential}: Whereas $V_{\rm AH}$ has a negative curvature around its origin, $V_{\rm CW}$ has a plateau due to the absence of the $|\Phi|^2$ term.
This difference can be seen in Fig.~\ref{fig: comparison of potentials}. Indeed, this fact entails a crucial difference between $V_{\rm AH}$ and $V_{\rm CW}$ in the dependence of energy composition of two strings separated at a finite distance $d$. We see this in Sec.~\ref{sec:two_string}.

A key quantity characterizing the features of the strings is the parameter $\beta$.
It is defined as the mass ratio between the gauge and scalar bosons
\begin{align}
\beta
&\equiv
\frac{m_\Phi^2}{m_A^2}
=
\left\{
\begin{array}{ll}
\displaystyle
\frac{4 \lambda_{\rm AH} v_\Phi^2}{2 g^2 v_\Phi^2}
=
\frac{2 \lambda_{\rm AH}}{g^2}
&~~:~~{\rm AH},
\\[0.5cm]
\displaystyle
\displaystyle
\frac{4 \lambda_{\rm CW} v_\Phi^2}{2 g^2 v_\Phi^2}
=
\frac{2 \lambda_{\rm CW}}{g^2}
&~~:~~{\rm CW}.
\end{array}
\right.
\label{eq: beta}
\end{align}
Note that, if the gauge boson $A_\mu$ dominantly contributes to the running in the CW case, $\lambda_{\rm CW} = 3 g^4 / 16 \pi^2$ gives $\beta = 3 g^2 / 8 \pi^2$ and hence perturbative calculations are not applicable in the $\beta \gtrsim 1$ regime.
However, as noted above, we regard $\lambda_{\rm CW}$ as a free parameter in order to accommodate the possibilities that other fields contribute to the running and determine the potential shape.

As mentioned in Sec.~\ref{sec:introduction}, in the AH case, $\beta$ is a parameter classifying type-I ($\beta<1$) and type-II ($\beta>1$) superconductors. In particular, for the critical coupling $\beta=1$ the BPS state is realized. In the CW case, however, such a classification is unclear at this point. We discuss it in Sec.~\ref{sec:two_string}.

\subsection{Comments on the use of Coleman-Weinberg potential}
\label{subsec:validity}

Before moving on to the analysis of vortex strings, we comment on the caveats of using the CW potential. The analysis of the ANO string has been done in the classical action \eqref{eq: starting action} with the quadratic-quartic potential \eqref{eq: SM potential}, while our attempt is made by considering the string dynamics with quantum-dressed potential $V_{\rm CW}$. However, the use of the CW potential alone in the analysis of topological defects may not be fully justified since the effective potential is merely the leading term in the derivative expansion of the effective action. In other words, quantum corrections not only deform the potential but also induce an infinite number of derivative operators which does not appear in the classical action. More specifically, we write schematically the effective action 
\begin{align}
\Gamma&= \int \df^4x\left [ \tilde V(\Phi_{\rm B}) + Z^{(0)}_\Phi|\p_\mu \Phi_{\rm B}|^2 +  Z^{(1)}_\Phi|\Phi_{\rm B}|^2|\p_\mu \Phi_{\rm B}|^2+ \cdots + Y^{(0)}_\Phi|\p^2 \Phi_{\rm B}|^2 + \cdots\right] \nn
&=\int \df^4x\left [V(\Phi) + |\p_\mu \Phi|^2 + \tilde Z^{(1)}_\Phi|\Phi|^2|\p_\mu \Phi|^2 + \cdots +\tilde Y^{(0)}_\Phi|\p^2 \Phi|^2+\cdots\right]\,,
\label{eq: schematic effective action}
\end{align}
where $\Phi_{\rm B}$ denotes the bare scalar field and the renormalized scalar field is defined as $\Phi=\sqrt{Z_\Phi^{(0)}}\Phi_{\rm B}$. Here $\tilde V(\Phi_B)$ is the effective potential before the field renormalization.
In general, the existence of higher derivative operators modifies the equation of motion for $\Phi$.
Our model-setup and analysis in this work correspond to the study within the local potential approximation~\cite{Hasenfratz:1985dm} such that in the effective action \eqref{eq: schematic effective action} we set $Z_\Phi^{(0)}=1$, $\tilde Z^{(i)}_\Phi=\tilde Y_\Phi^{(i)} =\cdots =0$ and $V(\Phi)$ is given by the CW potential~\eqref{eq: CW potential}. Although it is expected that higher derivative operators are subdominant and thus negligible in the low energy regime, it is difficult to completely guarantee the validity of this approximation.

Nevertheless, we would like to highlight through this work that energy barriers, which we will observe in the system of two CW-ANO strings, are a universal feature when the potential is flatter than quadratic around the origin. Indeed, this fact can be observed in other forms of the potential as we discuss in Appendix~\ref{app:universality}. Thus, the CW potential can at least be understood as one representative example of such potentials. We leave a complete analysis including other terms of the effective action for future work.

\subsection{Conversion to dimensionless quantities}

Throughout this paper, it is convenient to use dimensionless quantities since the dimensionality of the system is characterized by the single scale $v_\Phi$.
Here we define those more appropriately.
Let us start by rescaling the field variables,
\begin{align}
 A_\mu \to \frac{1}{g}A_\mu, \qquad \Phi \to \frac{1}{g} \Phi ,
\end{align}
leading to 
\begin{align}
S
&=
\frac{1}{g^2} \int \df^4 x
\left( |D_\mu \Phi|^2 - \frac{1}{4} F_{\mu \nu} F^{\mu \nu} -  V_\beta(\Phi) \right),
\label{eq: rescaled action}
\end{align}
with
\begin{align}
V_\beta(\Phi)
 &\equiv
\left\{
\begin{array}{ll}
\displaystyle
 \frac{\beta}{2} \Big( |\Phi|^2 - g^2 v_\Phi^2 \Big)^2
&~~:~~{\rm AH},
\\[0.5cm]
\displaystyle
 \frac{\beta}{2}
 \left(\ln \frac{|\Phi|^2}{g^2 v_\Phi^2} - \frac{1}{2} \right) |\Phi|^4
&~~:~~{\rm CW},
\end{array}
\right.
\end{align}
and $ D_\mu = \partial_\mu - i A_\mu $, in which $g$ does not appear.
The mass ratio $\beta$ is given in Eq.~\eqref{eq: beta} for the AH and CW cases.
Since $g$ and $v_\Phi$ always appears in the combination $g v_\Phi$, and since one dimensionful parameter can always be taken to be unity, we adopt the unit $g  v_\Phi = 1$.
It is equivalent to introduce the dimensionless variables (denoted by tilde)
\begin{align}
 &A_\mu = g v_\Phi \tilde A_\mu ,&
 &x^\mu = \tilde x^\mu / (g v_\Phi), &
 & \Phi = g  v_\Phi \tilde \Phi  ,
\end{align}
for which the action is given by
\begin{align}
S
&=
\frac{1}{g^2} \int \df^4 \tilde x
\left( |\tilde D_\mu \tilde \Phi|^2 - \frac{1}{4} \tilde F_{\mu \nu} \tilde F^{\mu \nu} - \tilde V(\tilde \Phi) \right)
\equiv \frac{1}{g^2}\tilde S,
\label{eq: dimensionless action}
\end{align}
with the covariant derivative $\tilde D_\mu=\tilde \p_\mu - i\tilde A_\mu$, the field strength $\tilde F_{\mu\nu}= \tilde \p_\mu \tilde A_\nu - \tilde \p_\nu \tilde A_\mu$, and the potential 
\begin{align}
\tilde V(\tilde \Phi)
 &\equiv
\left\{
\begin{array}{ll}
\displaystyle
\tilde V_\mathrm{AH} 
= \frac{\beta}{2} \Big( |\tilde \Phi|^2 - 1 \Big)^2
&~~:~~{\rm AH},
\\[0.5cm]
\displaystyle
\tilde V_\mathrm{CW}
= \frac{\beta}{2}
 \left(\ln |\tilde \Phi|^2 - \frac{1}{2} \right) |\tilde \Phi|^4
&~~:~~{\rm CW}.
\end{array}
\right.
\label{eq: rescaled potential}
\end{align}
Now we see that the action is written in terms of dimensionless quantities and contains no apparent scale. In this convention, it is clear that the dynamics of the theory only depends on the single dimensionless parameter $\beta$. One can easily translate all quantities in this dimensionless unit into those in the physical unit by multiplying $g v_\Phi $ with appropriate powers. The action \eqref{eq: dimensionless action} has the overall factor $1/g^2$, but it does not affect the string dynamics, and thus we use $\tilde S$ instead of the original $S$.
In the following sections, we always work in this convention and remove the tilde on dimensionless quantities for notational simplicity.

\section{Axisymmetric string solution}
\label{sec:axisymmetric}

In this section, we investigate axisymmetric ANO string solutions in the Abelian-Higgs model with the CW potential, and compare them with the solutions for the conventional quadratic-quartic potential.
In particular, we highlight how the energy of the solutions depend on $\beta$ (the ratio between scalar and gauge boson masses).

\subsection{ANO string solution}
\label{sec:ANO string solution}

As shown by Nielsen and Olesen in Ref.~\cite{Nielsen:1973cs}, the Abelian Higgs model \eqref{eq: starting action}, in general, has a vortex string solution as a non-trivial (classical) solution to its equation of motion.
This is ensured whenever the potential $V(\Phi)$ has $U(1)$-breaking vacua (i.e., the vacua characterized by a non-trivial first homotopy group). Thus the existence does not depend on the detailed shape of the potential $V(\Phi)$.

To find the solutions, we start by assuming static and axially symmetric configurations and then parametrizing the fields $\Phi$ and $A_\mu=(A_t, A_r,A_\theta,A_z)$ as\footnote{
Note $A_x = - A_\theta \sin \theta / r$ and $A_y = A_\theta \cos \theta / r$.
}
\begin{align}
\Phi
&= f (r) e^{i n \theta},
\qquad
A_\theta = n a(r),
\qquad
A_t = A_z = A_r= 0.
\label{eq: stationary configurations}
\end{align}
Here $n$ is the winding number being integers and $r=\sqrt{x^2+y^2}$ is the (dimensionless) radius on the $xy$-plane. 
For their regularity and finiteness of the energy, the profile functions $f(r)$ and $a(r)$ satisfy the boundary conditions
\begin{align}
    f(0)=a(0)=0, \qquad f(\infty)=a(\infty)=1 .
\label{eq: boundary condition}
\end{align}
Inserting Eq.~\eqref{eq: stationary configurations} into the energy per unit length (i.e. tension) yields
\begin{align}
T \equiv \frac{\df E}{\df z}
&=
\int r \df r \, \df \theta
\left[
\left( \frac{\df f}{\df r} \right)^2
+ \frac{n^2}{2 r^2} \left( \frac{\df a}{\df r} \right)^2
+ \frac{n^2}{r^2} f^2 (1 - a)^2
+ V (f)
\right].
\label{eq: rescaled energy}
\end{align}
The rescaled potential $V$ defined in Eq.~\eqref{eq: rescaled potential} is given respectively by
\begin{align}
V (f)
&=
\left\{
\begin{array}{ll}
\displaystyle
\frac{\beta}{2} (f^2 - 1)^2
&~~:~~{\rm AH},
\\[0.5cm]
\displaystyle
\frac{\beta}{2} f^4 \left( \ln (f^2) - \frac{1}{2} \right)
&~~:~~{\rm CW}.
\end{array}
\right.
\label{eq: dimensionless potentials}
\end{align}
Here the coefficient of the potentials is given in terms of the mass ratio $\beta$ defined in Eq.~\eqref{eq: beta}.
We see that only $\beta$ is a free parameter of the system.
Varying the tension \eqref{eq: rescaled energy} with respect to $f$ and $a$, their equations of motion are found to be
\begin{align}
&f'' + \frac{1}{r} f' -\frac{n^2 (1-a)^2}{r^2}f  - \frac{1}{2}\frac{\partial V}{\partial f}=0,
\label{eq: ano eom for f}
\\[1ex]
&a'' -\frac{1}{r}a' +2 (1-a)f^2=0,
\label{eq: ano eom for a}
\end{align}
respectively.
Here and hereafter the prime denotes the derivative with respect to $r$, e.g. $f'=\df f / \df r$.

We read off the energy density for the stationary configurations \eqref{eq: stationary configurations} as
\begin{align}
{\cal E} (r)
=
(f')^2 + \frac{n^2}{2 r^2} (a')^2 + \frac{n^2}{r^2} f^2 (1 - a)^2 + V (f).
\label{eq: energy composition in a string}
\end{align}
In particular,  for $V_{\rm AH}$, the tension \eqref{eq: rescaled energy} can be rewritten as
\begin{align}
T &= 2 \pi \int_0^\infty \df r\,r\, {\cal E} (r) \nn
&= 2\pi |n| \nn
&\quad
+ 2\pi\int_0^\infty \df r \, r \left[ \left( f' + |n| \frac{a - 1}{r}f \right)^2 + \frac{n^2}{2r^2}\left( a' + \frac{r}{|n|} (f^2 -1) \right)^2 +\frac{1}{2}(\beta-1) (f^2 - 1)^2 \right].
\label{eq: tension in complete squared}
\end{align}
The first and second terms in the integrand are in squared forms and thus always give positive values.
Therefore, the tension is bounded from below as
\begin{align}
T \geq 2\pi |n| +2\pi\int_0^\infty \df r\, r \left[  \frac{1}{2}(\beta-1) (f^2 - 1)^2 \right].
\label{eq: bound for energy}
\end{align}
Moreover, if $\beta\geq 1$, the second term in Eq.~\eqref{eq: bound for energy} also becomes non-negative, so that one has the Bogomol'nyi bound; $T\geq 2\pi|n|$.  In particular, for $\beta=1$ the last term in Eq.~\eqref{eq: tension in complete squared} vanishes, and the equations of motion \eqref{eq: ano eom for f} and \eqref{eq: ano eom for a} can be rewritten as the first-order Bogomol'nyi equations~\cite{Bogomolny:1975de} in terms of $f$ and $a$:
\begin{align}
&
f' + |n| \frac{a - 1}{r}f = 0,
\qquad
a' + \frac{r}{|n|} (f^2 - 1) = 0,
\label{eq: Bogomol'nyi equations}
\end{align}
for which the tension is given by the Bogomol'nyi limit
\begin{align}
T=2\pi|n| \qquad (\beta=1).
\label{eq: Bogomol'nyi limit}
\end{align}
Hence, the vortex becomes stable in sense that its tension takes the lowest value of the energy bound. This is the BPS state.
In the case of $V_{\rm AH}$, the stability of the vortex strings can be understood analytically in terms of $\beta$, but this is not possible for the CW potential $V_{\rm CW}$.
Thus the stability analysis of the latter should instead rely on numerical methods.
Clarifying the stability of vortex strings for the CW type potential is one of the main purposes in this work, and is done in Sec.~\ref{sec:two_string}.


We close this subsection by mentioning the behavior of $f$ and $a$ as functions of $r$. Since exact solutions to Eqs.~\eqref{eq: ano eom for f} and \eqref{eq: ano eom for a} do not exist even in the case of the AH potential, numerical methods are necessary to obtain the full solutions.
This is discussed in the next subsection. Instead, we here explore the asymptotic behaviors of $f$ and $a$ in an analytic way. 
In the limit $r \to \infty$, the functions $f$ and $a$ are sufficiently close to the vacuum values, and hence it is convenient to rewrite the equations of motion to the leading order in $\delta f \equiv 1-f \ll 1$ and $\delta a \equiv 1-a \ll 1$ as
\begin{align}
&\delta f'' + \frac{1}{r} \delta f' - 2\beta \delta f = \mathcal{O}((\delta f)^2, (\delta a)^2) ,
\label{eq: linear eom for f}
\\[1ex]
&\delta a'' -\frac{1}{r} \delta a' - 2 \delta a= \mathcal{O}((\delta f)^2, (\delta a)^2) ,
\label{eq: linear eom for a} 
\end{align}
from which it is clear that $\delta f$ and $\delta a$ behave as 
\begin{equation}
    \delta f \propto r^{-1/2}\, \exp \left[- \sqrt{2\beta} r \right] , \qquad \delta a \propto r^{1/2}\, \exp \left[- \sqrt{2} r\right] , 
    \label{eq: asymptotic}
\end{equation}
for $r \to \infty$.
Note that these expressions apply only for $\beta < 4$.
For $\beta > 4$ (i.e. $m_H > 2 m_A$), the nonlinear contribution $(\delta a)^2$ is larger than the linear contribution $\delta f$ and thus cannot be neglected~\cite{Perivolaropoulos:1993uj}.
In this case, the linearized EOMs for $\delta f$ should be modified to be
\begin{align}
&\delta f'' + \frac{1}{r} \delta f' - \frac{1}{r^2} (\delta a)^2 - 2\beta \delta f = \mathcal{O}((\delta f)^2, (\delta a)^3) ,
\label{eq: linear eom2 for f}
\end{align}
where the third term is the source term for $\delta f$ behaving as $r^{-1}\,e^{- 2 \sqrt{2} r}$, and it leads to the asymptotic behavior
\begin{equation}
    \delta f \propto r^{1/2}\, \exp \left[- 2\sqrt{2} r \right] .
\end{equation}
On the other hand, $\delta a $ does not change from Eq.~\eqref{eq: asymptotic}.
These results are independent of the detailed shape of the potential $V(f)$, and thus they hold both for the AH and CW cases, because the nonlinear terms with respect to $\delta f$ are always negligible.
We check this fact numerically in the next subsection.

\subsection{Numerical result}
\label{sec:Numerical result for one string}

We discuss the behavior of $f$ and $a$ by solving the equations of motion \eqref{eq: ano eom for f} and \eqref{eq: ano eom for a} numerically.
However, it is technically problematic to directly deal with Eqs.~\eqref{eq: ano eom for f} and \eqref{eq: ano eom for a} due to numerical fine-tuning required.
Instead, we here use the relaxation method (a.k.a. the gradient flow method) in order to obtain the static configuration with the minimum energy numerically.
We introduce a fictitious time $\tau$ called the flow time instead of the real time $t$, and promote the profile functions $f(r)$ and $a(r)$ to $\tau$-dependent functions, $f(r,\tau)$ and $a(r,\tau)$.
We evolve them by the following differential equations (flow equations): 
\begin{align}
f'' + \frac{1}{r} f' -\frac{n^2 (1-a)^2}{r^2}f - \frac{1}{2}\frac{\partial V}{\partial f}
&=
\partial_\tau f, \label{eq:diff-f}
\\
a'' -\frac{1}{r}a' +2 (1-a)f^2
&=
\partial_\tau a, \label{eq:diff-a}
\end{align}
starting from some appropriate functions satisfying the boundary conditions \eqref{eq: boundary condition} as the initial configuration at $\tau=0$. For instance, we set $f(r,0)=\tanh(r)$ and $a(r,0)=\tanh^2(r)$.
If the $\tau$-evolution converges, $\partial_\tau f=\partial_\tau a=0$, the converged profile functions are nothing but the static solution of the original equations of motion.

\begin{figure}
\begin{center}
\includegraphics[width=0.48\columnwidth]{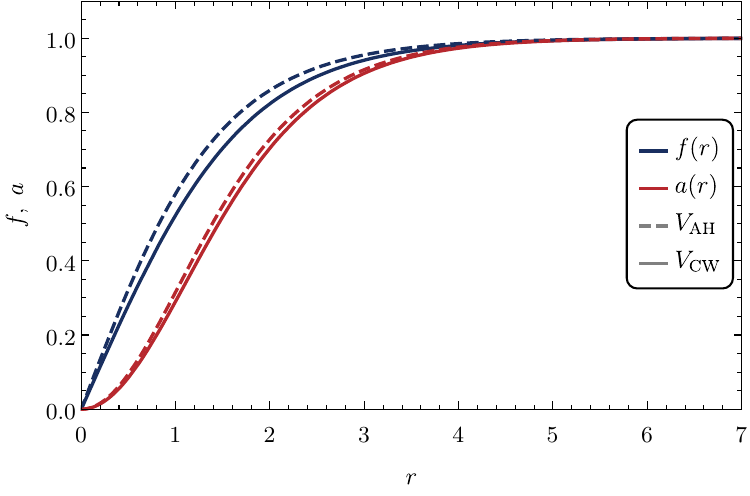}
\hspace{2ex}
\includegraphics[width=0.48\columnwidth]{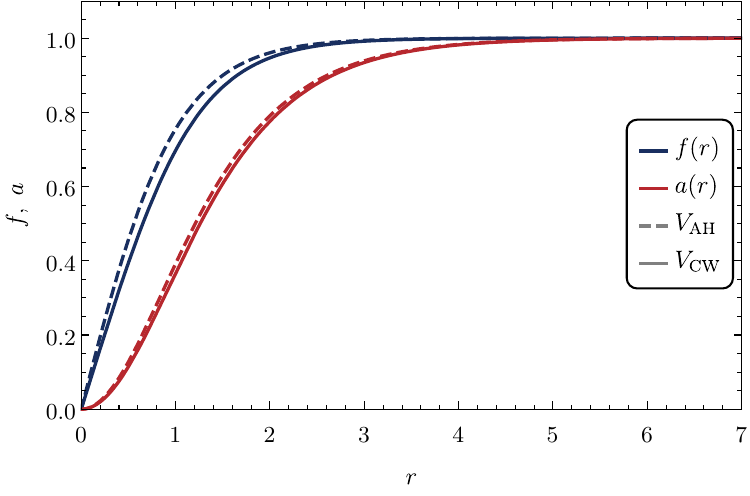}
\caption{\small ANO vortex string solutions to Eqs.~\eqref{eq: ano eom for f} and \eqref{eq: ano eom for a} for $n=1$ with the potentials \eqref{eq: dimensionless potentials}. We set $\beta=0.5$ (left) and $\beta=1.5$ (right).
}
\label{fig: example of ano solution}
\end{center}
\end{figure}

\begin{figure}
\begin{center}
\includegraphics[width=\columnwidth]{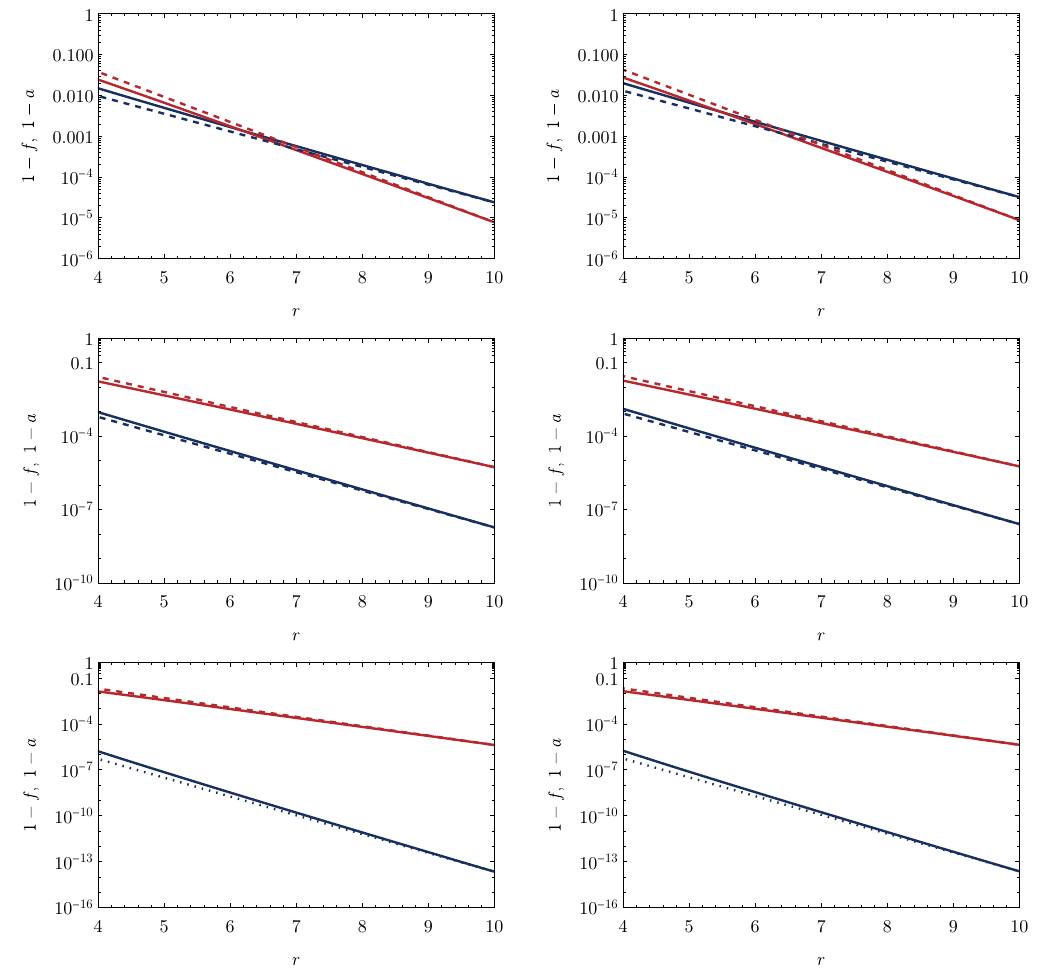}
\caption{\small
Asymptotic behavior of the Higgs field $1 - f$ (blue) and the gauge field $1 - a$ (red) for the AH (left) and CW (right) strings.
The parameter value is $\beta = 0.5$, $1.5$, and $8$ from top to bottom.
The dashed lines are $\propto e^{- \sqrt{2 \beta} r}$ (blue) and $\propto e^{- \sqrt{2} r}$ (red).
As well known, for $\beta > 4$ (i.e. $m_H > 2 m_A$), the asymptotic behavior of the Higgs field deviates from $\propto e^{- \sqrt{2 \beta} r}$ but rather follows $\propto e^{- 2 \sqrt{2} r}$, as shown in the blue dotted lines in the bottom panels.
}
\label{fig:asymptotic}
\end{center}
\end{figure}

In Fig.~\ref{fig: example of ano solution}, we show numerical solutions for $f(r)$ and $a(r)$ with $n=1$ for both $V_{\rm AH}$ and $V_{\rm CW}$ given in Eq.~\eqref{eq: dimensionless potentials} with $\beta=0.5$ and $1.5$. We observe no drastic difference between the ANO vortex solutions with $V_{\rm AH}$ and $V_{\rm CW}$. 
Fig.~\ref{fig:asymptotic} shows the asymptotic behavior of the AH-ANO and CW-ANO strings obtained by the numerical calculations for $\beta = 0.5$, $1.5$, and $8$.
The blue and red lines represent $\delta f \equiv 1 - f$ and $\delta a \equiv 1 - a$, respectively.
These quantities follow the analytic prediction $\propto e^{- \sqrt{2 \beta} r}$ and $\propto e^{- \sqrt{2} r}$ for large $r$ for $\beta=0.5$ and $1.5$, as seen from the top and middle panels.
For $\beta=8$, the asymptotic behavior of the Higgs field deviates from $\propto e^{- \sqrt{2 \beta} r}$ but rather follows $\propto e^{- 2 \sqrt{2} r}$, as seen from the bottom panels.
Thus they are consistent with the analytic prediction obtained in Eq.~\eqref{eq: asymptotic}.

\begin{figure}
\begin{center}
\includegraphics[width=\columnwidth]{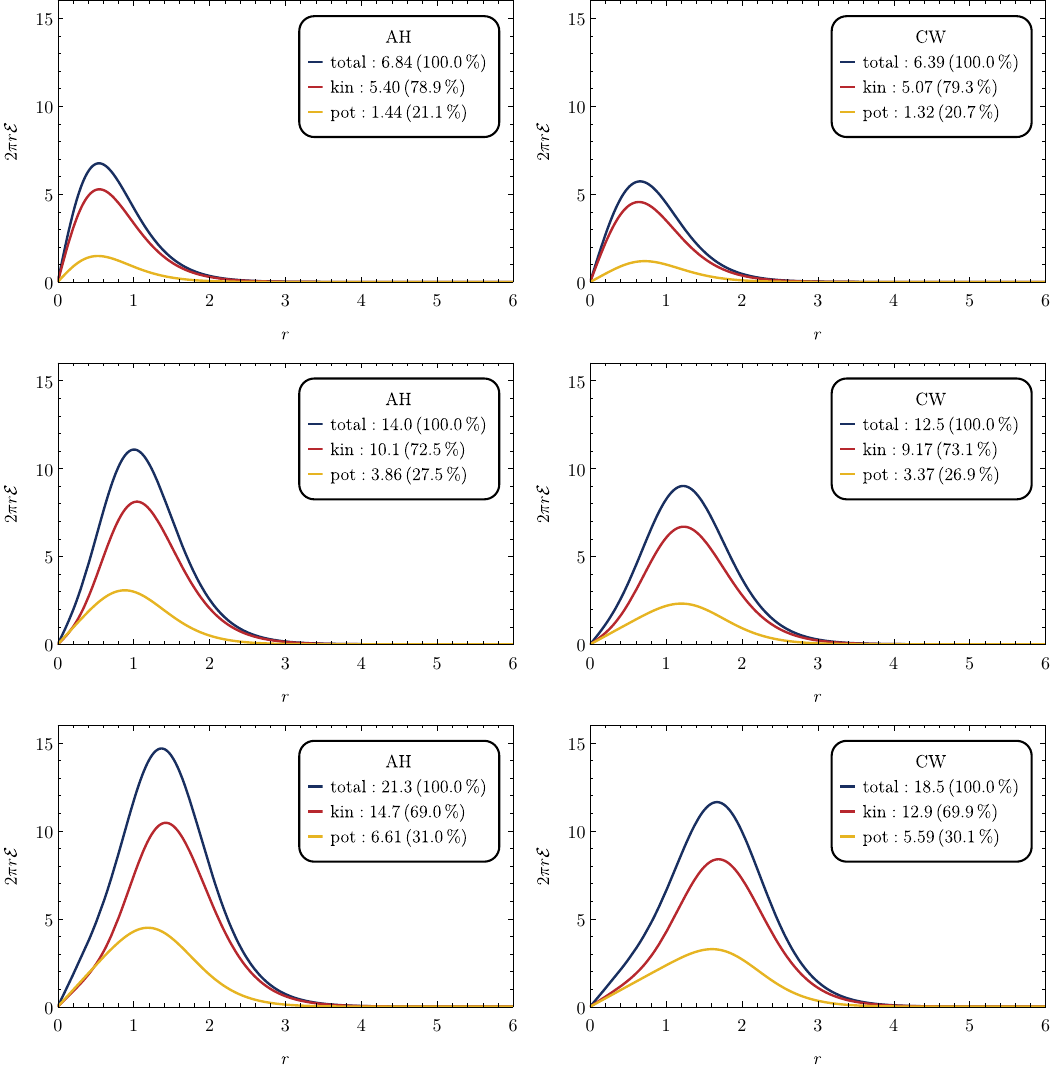}
\caption{\small
Energy composition of the strings with $n = 1$, $2$, and $3$ from top to bottom for the AH (left) and CW (right) potentials with $\beta = 1.5$.
}
\label{fig:d_0_composition}
\end{center}
\end{figure}

Fig.~\ref{fig:d_0_composition} shows the energy composition \eqref{eq: energy composition in a string} of the AH-ANO and CW-ANO strings with winding number $n = 1$, $2$, and $3$ from top to bottom for $\beta = 1.5$.
Here we define the contribution from $V$ to the total energy as the potential contribution, and the rest as the kinetic
\begin{align}
T_V
&\equiv
\int r \df r \, \df \theta~V,
\qquad
T_K
\equiv
T - T_V.
\end{align}
We see that the fraction of each contribution does not differ much between the two cases, while their values themselves are slightly smaller in the CW-ANO string than those of the AH-ANO string.
We also see that the energy peak is located slightly outward for the CW-ANO string, reflecting the flat structure of the potential.

\begin{figure}
\begin{center}
\includegraphics[width=\columnwidth]{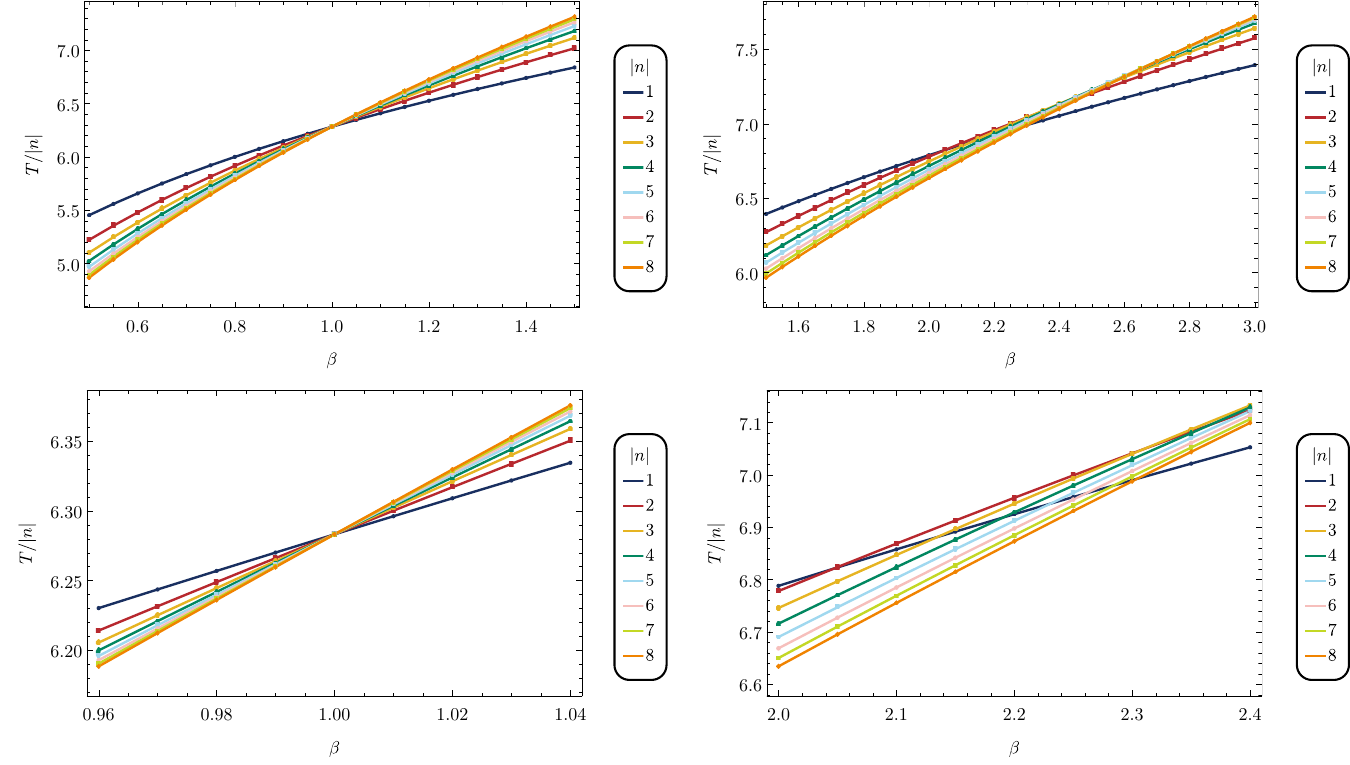}
\caption{\small
Energy of the string with different winding number $n$ for the AH (left) and CW (right) potentials.
The bottom panels are zoom-in of the top panels.
}
\label{fig:d_0_beta-E}
\end{center}
\end{figure}

We finally plot the total energy of the string for different values of $\beta$ and for different values of the winding number in Fig.~\ref{fig:d_0_beta-E}.
The left and right panels are for the AH and CW potentials, respectively, and the bottom row is a zoom-in of the top row.
For the AH potential, all the lines cross at $\beta = 1$ that corresponds to the BPS limit (critical coupling).
Besides, the value of $T/|n|$ at $\beta=1$ takes $2\pi \simeq 6.28$.
At this parameter point, the energy does not depend on the number of strings overlain. This fact indeed agrees with Eq.~\eqref{eq: Bogomol'nyi limit}.
For the CW potential, in contrast, the lines do not cross at a single point.
Therefore, it seems that there is no apparent BPS state for the CW potential.
Comparing the lines with $n = 1$ and $n = 2$, we infer naively that the force between two strings with winding number $n = 1$ is attractive for $\beta \lesssim 2.1$, while it is repulsive for $\beta \gtrsim 2.1$ (see also Fig.~\ref{fig:stable-metastable}).
However, as for whether the actual force acting between the two strings is attractive or repulsive, nontrivial dependence may arise depending on the distance $d$.
We discuss this point in detail in the next section to elucidate the interaction feature of the strings.


\section{Interaction potential for two string system}
\label{sec:two_string}

In this section we investigate the interaction between two parallel CW-ANO strings located at the interstring distance $d$.
The static energy $E$ (per unit length) of this system for given $d$ is regarded as an effective interaction potential for the strings, which is useful for discussing the stability of the system.

\subsection{Two-string system}
\label{subsec:Setup}

The underlying model action is the same as Eq.~\eqref{eq: starting action} (or its dimensionless version \eqref{eq: dimensionless action}).
We consider a system with two parallel CW-ANO strings extending in the $z$ direction.
Thanks to the translational invariance in $z$, it is sufficient to describe the two strings on the orthogonal plane using static ansatz depending on the two-dimensional $(x, y)$ coordinate
\begin{align}
    \Phi= \Phi_1(x,y) + i \Phi_2(x,y) , \qquad A_\mu= (0,A_1(x,y), A_2(x,y),0).
\end{align}
Here $\Phi_1$ and $\Phi_2$ are real functions.
In this ansatz, however, there is an issue of the gauge redundancy. This often leads to technical problems in numerical computations such as convergence.
Thus, we fix the gauge as the Coulomb gauge by adding the following gauge fixing action
\begin{equation}
S_\text{g.f.} =  
-\frac{1}{2} \int \df^4x \, (\partial_i A^i)^2,
\end{equation}
giving the gauge-fixed action
\begin{align}
S
&=
\int \df^4x
\left[
|D_\mu \Phi|^2 - \frac{1}{4} F_{\mu \nu} F^{\mu \nu} - V(\Phi) - \frac{1}{2} (\partial_i A^i)^2
\right].
\end{align}
Using this ansatz, the (gauge-fixed) tension reads off as 
\begin{align}
T=\frac{\df E}{\df z} = \int \df x\,\df y\, {\cal E}(x,y),
\end{align}
with the energy density
\begin{align}
{\cal E}
&=
|\partial_i \Phi|^2 + \frac{1}{4} (\partial_i A_j - \partial_j A_i)^2 + \frac{1}{2} (\partial_i A_i)^2 + A_i^2 |\Phi|^2 + i A_i (\Phi^* \partial_i \Phi - \Phi \partial_i \Phi^*) + V(\Phi),
\label{eq: energy 2d}
\end{align}
where the index $i$ runs $1,2$.
The potential $V$ is defined by Eq.~\eqref{eq: rescaled potential} with the tildes removed.
To calculate the interaction potential for given $d$, we put two CW-ANO strings at $(x,y)=(\pm d/2,0)$, and  minimize the energy \eqref{eq: energy 2d} with the positions of the string cores fixed.
The minimized energy value is the interaction potential energy at $d$.
By performing this for various $d$, we obtain the structure of the interaction potential.

As in the axisymmetric case, we rely on the numerical calculation as it is difficult to perform the minimization procedure analytically.
For the same reason as mentioned in Sec.~\ref{sec:Numerical result for one string}, we promote the functions $\Phi_1, \Phi_2, A_1$ and $A_2$ to $\tau$-dependent ones and achieve the minimization procedure by solving the diffusion equation
\begin{align}
-\frac{\delta \mathcal{E}}{\delta X} 
&=\partial_\tau X,
\label{eq: diff eq 2d}
\end{align}
with $X$ denoting the functions $X=\Phi_1, \Phi_2, A_1$, and $A_2$,
until this ``time evolution'' sufficiently converges. Specifically, the diffusion equation \eqref{eq: diff eq 2d} reads
\begin{align}
\partial_i \partial_i \Phi_1 - A_i^2 \Phi_1 + 2 A_i \partial _i \Phi_2 - \frac{\partial V}{\partial \Phi_1}
&=\partial_\tau \Phi_1, 
\label{eq: 1. diffusion equations}
\\
\partial_i \partial_i \Phi_2 - A_i^2 \Phi_2 - 2 A_i \partial _i \Phi_1 - \frac{\partial V}{\partial \Phi_2} 
&=\partial_\tau \Phi_2,
\\
\partial_i \partial_i A_1 - 2(\Phi_1^2+\Phi_2^2) A_1 + 2 \left(\Phi_1 \partial_1 \Phi_2  - \Phi_2 \partial_1 \Phi_1 \right) 
&=\partial_\tau A_1,
\\
\partial_i \partial_i A_2 - 2(\Phi_1^2+\Phi_2^2) A_2 + 2\left(\Phi_1\partial_2 \Phi_2  - \Phi_2 \partial_2 \Phi_1 \right) 
&=\partial_\tau A_2.
\label{eq: 4. diffusion equations}
\end{align}
In the following subsections, we present the numerical solutions to these equations and discuss the properties of the two vortex strings.

\subsection{Numerical results}
\label{subsec:numerical}

We start by numerically solving Eqs.~\eqref{eq: 1. diffusion equations}--\eqref{eq: 4. diffusion equations}.
In order to fix the string cores, we impose $\Phi_1 (\tau, x, y) = 0$ at the position of the cores at every step of the time evolution.
We take the box size $L = 30$ with the grid size $\Delta L = 0.05$ for both $x$ and $y$ directions, and evolve the diffusion equations from $\tau = 0$ to $\tau = 15$ with the time step $\Delta \tau = 0.0005$.
In the following we plot only part of the full box for visibility.

\subsubsection{Field configurations}

\begin{figure}
\begin{center}
\includegraphics[width=\columnwidth]{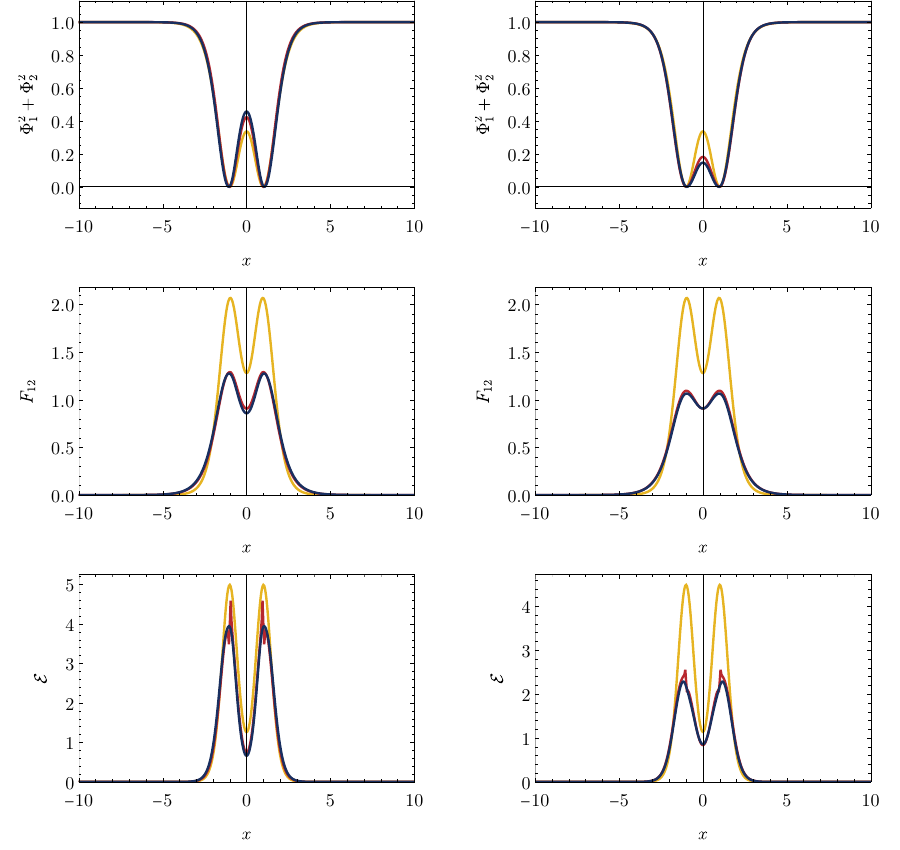}
\caption{\small
Field configurations along $y = 0$ for the AH-ANO string (left) and for the CW-ANO string (right) for $\beta = 2$.
Different lines are for $t = 0$ (yellow), $2 $ (red), and $t = 15 $ (blue).
For the spikes in the bottom panels, see footnote \ref{fn:spike}.
}
\label{fig:configuration_slice}
\end{center}
\end{figure}
\begin{figure}
\begin{center}
\includegraphics[width=\columnwidth]{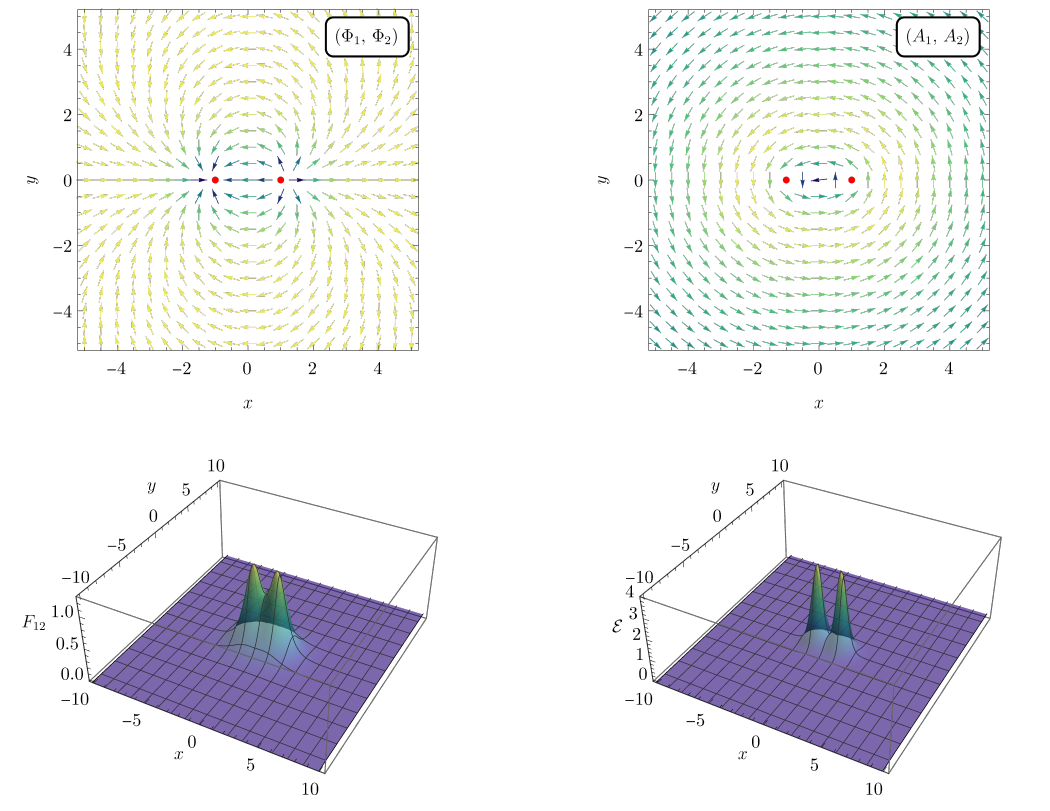}
\caption{\small
Field configurations for the AH-ANO string for $\beta = 2$ and $d = 2$ with winding numbers $(n_L, n_R) = (1, 1)$.
The top panels are $(\Phi_1, \Phi_2)$ (top-left) and $(A_1, A_2)$ (top-right),
while the bottom panels are $F_{1 2}$ (bottom-left) and ${\cal E}$ (bottom-right).
The red dots in the top panels are the position of the string cores.
The color in the top panels corresponds to the norm of the vectors, $(\Phi_1,\Phi_2)$ and $(A_1,A_2)$.
}
\label{fig:configuration_AH}
\end{center}
\end{figure}
\begin{figure}
\begin{center}
\includegraphics[width=\columnwidth]{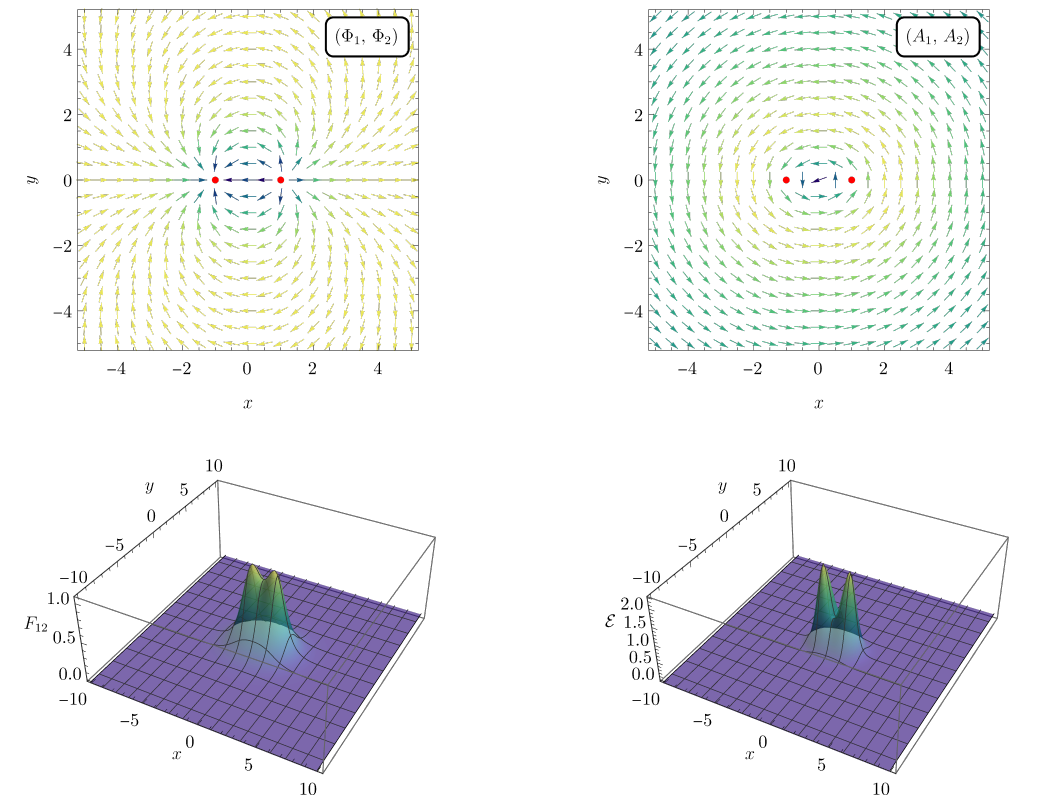}
\caption{\small
Field configurations for the CW-ANO string for $\beta = 2$ and $d = 2$ with winding numbers $(n_L, n_R) = (1, 1)$.
The top panels are $(\Phi_1, \Phi_2)$ (top-left) and $(A_1, A_2)$ (top-right),
while the bottom panels are $F_{1 2}$ (bottom-left) and ${\cal E}$ (bottom-right).
The red dots in the top panels are the position of the string cores.
The color in the top panels corresponds to the norm of the vectors, $(\Phi_1,\Phi_2)$ and $(A_1,A_2)$.
}
\label{fig:configuration_CW}
\end{center}
\end{figure}

We first see the field configurations.
Fig.~\ref{fig:configuration_slice} shows the $y = 0$ slice of the field configurations for $\beta = 2$ and $d = 2$ for the AH (left) and CW (right) strings.
The yellow, red, and blue lines are time slices at $t = 0 $, $2$, and $15 $, respectively.\footnote{
\label{fn:spike}
Due to $\Phi_1 = 0$ we impose at the string cores, the energy density develops small spikes at the cores (see the red lines in the bottom panels of Fig.~\ref{fig:configuration_slice}).
Although these spikes are negligible in the total energy, we remove them at the end of simulation $t = 15$ by further evolving the system by $20$ steps without imposing $\Phi_1 = 0$.
We checked that the effect of this procedure is negligible both on the string locations and on the total energy of the system.
In Fig.~\ref{fig:configuration_slice}, the final time slice (blue lines) shows the field configurations after this procedure.
}
The two peaks in the flux density and energy density correspond to the string cores, $(x, y) = (\pm 1 , 0)$, and we see that $\Phi_1^2 + \Phi_2^2$ takes zero at these points.
Each string has the winding number unity, so that the total winding number on the $xy$-plane is $n=2$.

One clear difference between the AH and CW cases is the height of the flux density and energy density.
We already observed this behavior in Sec.~\ref{sec:axisymmetric}: while the energy composition is not much different between the two cases, the energy density itself for the same value of $\beta$ is smaller for the CW-ANO string due to the flat structure of the potential around the origin.

Figs.~\ref{fig:configuration_AH} and \ref{fig:configuration_CW} are two-dimensional field configurations at $t = 15 $ for the same parameter point.
The phase of the complex scalar field $\Phi = \Phi_1 + i \Phi_2$ rotates twice along the circle with $r = \infty$, because the total winding number of the system is two.
Both figures do not have much difference except for the height of the peaks in the flux and energy densities, as mentioned above.

\subsubsection{Attractive/repulsive force between strings}

\begin{figure}
\begin{center}
\includegraphics[width=\columnwidth]{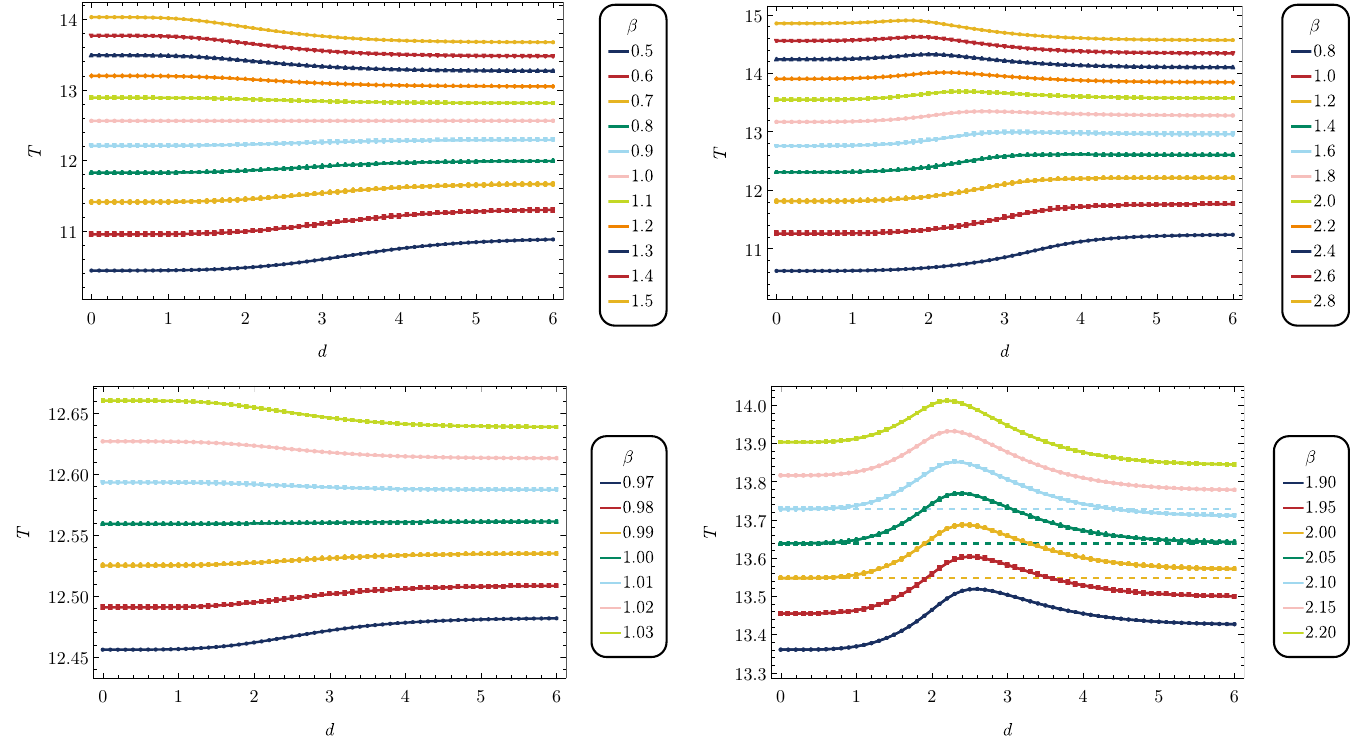}
\caption{\small
Distance dependence of the energy of the two-string system with the AH (left) and CW (right) potentials.
}
\label{fig:d>0_beta-E}
\end{center}
\end{figure}

We perform above numerical analysis for different values of $\beta$ and $d$ to construct the interaction potential for the two-string system with the winding number of the left and right strings being $(n_L,n_R)=(1,1)$ (i.e. $n=2$ in total).

Fig.~\ref{fig:d>0_beta-E} shows the dependence of the total energy per unit length (i.e. tension) on the interstring distance $d$.
The top row is a broad scan of $\beta$ for the AH (left) and CW (right) potentials, while the bottom row shows a zoom-in of the top row around the parameter values where the tension becomes almost the same at $d = 0$ and $d \to \infty$.
As seen from the left panels, the lines do not develop any nontrivial structure in the AH case.
In particular, the energy minimum appears at $d = 0$ ($d \to \infty$) for $\beta < 1$ ($\beta > 1$), which reflects the well-known fact that the AH-ANO string with winding number $n=2$ is stable for $\beta < 1$ while it is unstable and breaks up into two strings with $n = 1$ for $\beta > 1$.

In contrast, as shown in the right panels of Fig.~\ref{fig:d>0_beta-E}, the CW case develops an energy barrier for some range of $\beta$.
This energy barrier is most pronounced around $\beta \simeq 2$.
Due to this barrier, $d = 0$ is either a local or absolute minimum for this range of $\beta$.
Comparing the asymptotic values of the energy at $d = 0$ and $d \to \infty$, we find that $d = 0$ is an absolute minimum when $\beta$ is smaller than a critical value $\beta_c$ (see $\beta = 2.00, 1.95, 1.90$ in the right-bottom panel) 
while it becomes a local (not global) minimum when $\beta$ is larger than $\beta_c$ (see $\beta = 2.10, 2.15, 2.20$).
This means that CW-ANO strings are metastable once $\beta$ exceeds $\beta_c$.
While the energy barrier prevents the metastable CW-ANO string with $n = 2$ from breaking up classically, quantum effects allow for it.
Numerically we find that this stable-metastable transition occurs around $\beta_c \simeq 2.07$, see the top panel of Fig.~\ref{fig:stable-metastable}.

Note that, around $\beta \simeq 1$, the asymptotic behavior at large $d$ for the CW-ANO string is almost flat and hence the existence of the barrier is difficult to read off from the figure. 
This is because the asymptotic behavior is a superposition of two effects: one is the energy barrier, and the other is the mild exponentially decaying tail.
The latter is studied in the previous asymptotic analysis for axisymmetric strings.
What we find there is that the coefficient of the exponential is positive (negative) for $\beta < 1$ ($\beta > 1$), independently of the potential shape.
Thus we expect that the CW-ANO string develops the barrier for $\beta$ slightly larger than unity.
On the other hand, it is numerically hard to see if there is an upper bound on $\beta$ that develops the barrier.

Finally, we highlight the difference between the AH and CW cases from another viewpoint.
In the conventional AH-ANO string with the quadratic-quartic potential, $\beta = 1$ is the only value at which the system shows a clear transition between different regimes.
What we find here is that it is not always the case for more general potentials: For the CW-ANO string, $\beta = 1$ is still the transition between the attractive/repulsive regimes for two strings far separated, whereas another critical value $\beta = \beta_c$ comes into the game once (meta)stability is concerned.
Therefore, two-string systems with generic potentials may have richer phase structures and lead to richer phenomena than previously thought.

\subsection{Larger winding numbers}

\begin{figure}
\begin{center}
\includegraphics[width=\columnwidth]{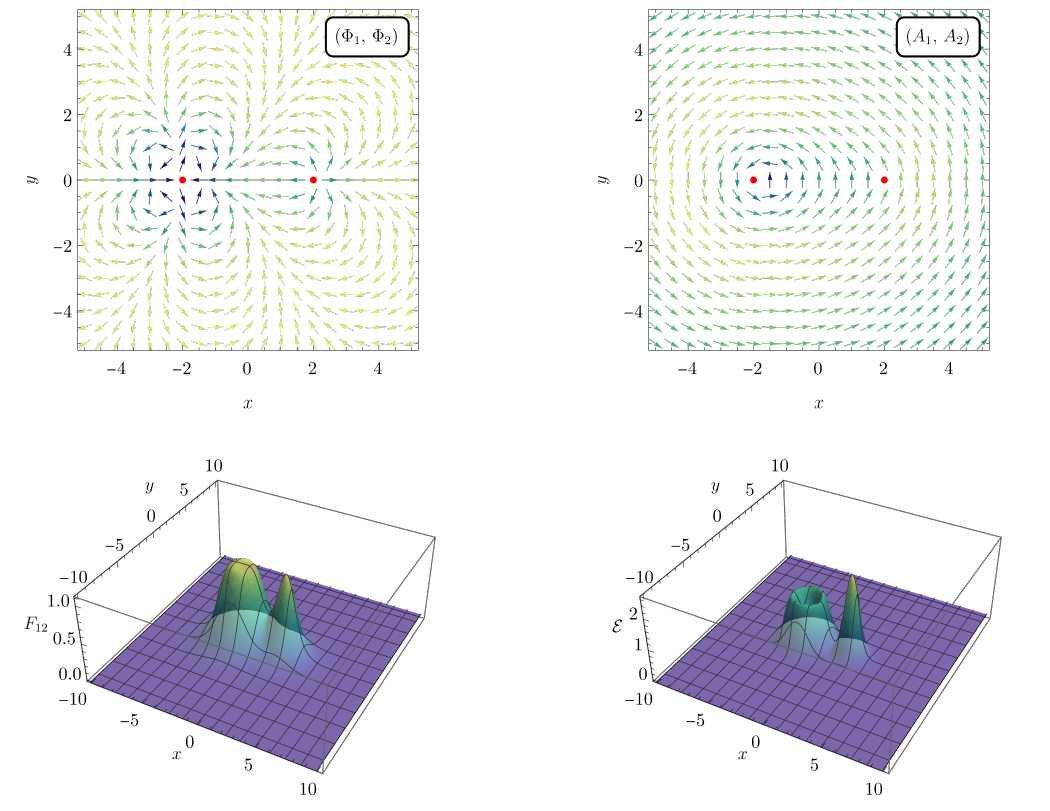}
\caption{\small
Field configurations for the CW-ANO string for $\beta = 2$ and $d = 4$ with winding numbers $(n_L, n_R) = (3, 1)$.
The top panels are $(\Phi_1, \Phi_2)$ (top-left) and $(A_1, A_2)$ (top-right),
while the bottom panels are $F_{1 2}$ (bottom-left) and ${\cal E}$ (bottom-right).
The red dots in the top panels are the position of the string cores.
The color in the top panels corresponds to the norm of the vectors, $(\Phi_1,\Phi_2)$ and $(A_1,A_2)$.
}
\label{fig:configuration_CW_largen}
\end{center}
\end{figure}
\begin{figure}
\begin{center}
\includegraphics[width=\columnwidth]{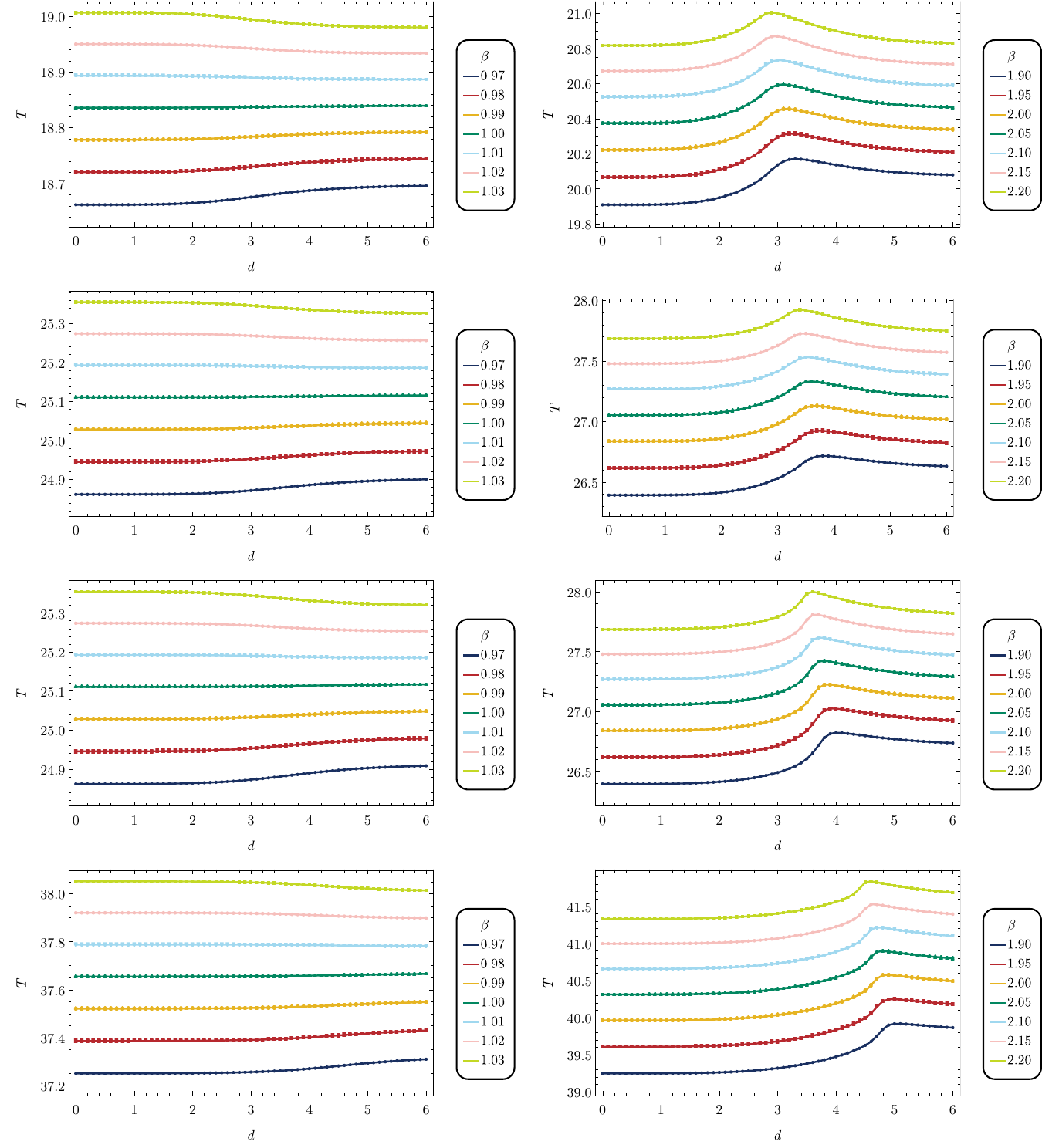}
\caption{\small
Distance dependence of the energy for larger winding numbers for the AH (left) and CW (right) strings.
The winding number is $(n_L, n_R) = (2, 1)$, $(3, 1)$, $(2, 2)$, and $(3,3)$ from top to bottom.
}
\label{fig:largen}
\end{center}
\end{figure}
\begin{figure}
\begin{center}
\includegraphics[width=\columnwidth]{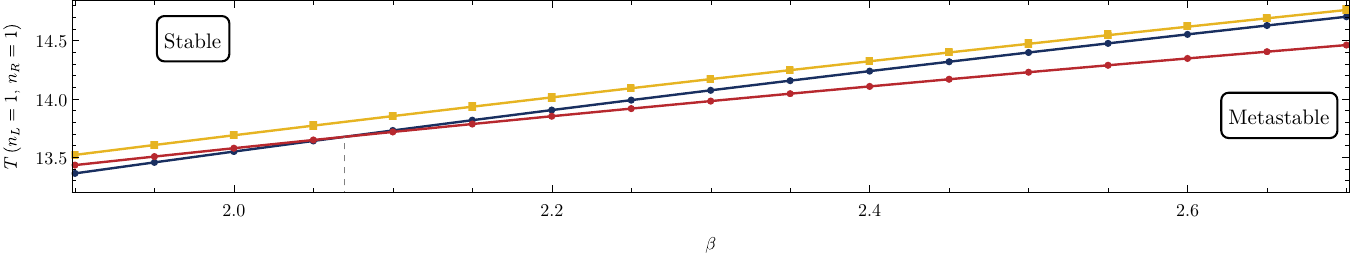}
\vskip 0.2cm
\includegraphics[width=\columnwidth]{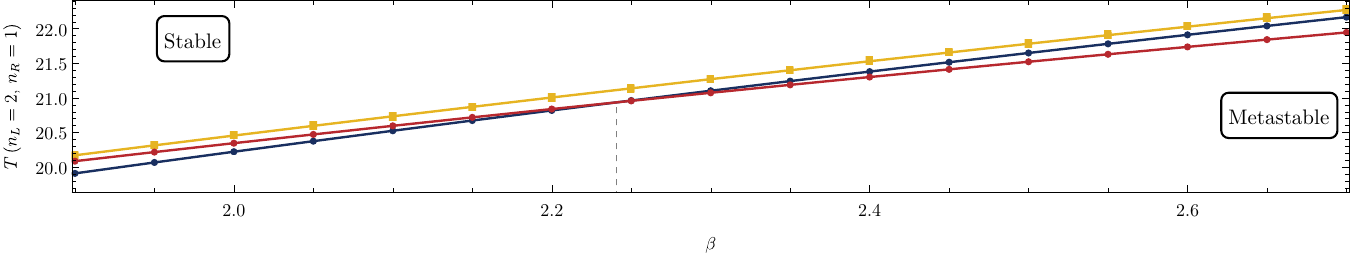}
\vskip 0.2cm
\includegraphics[width=\columnwidth]{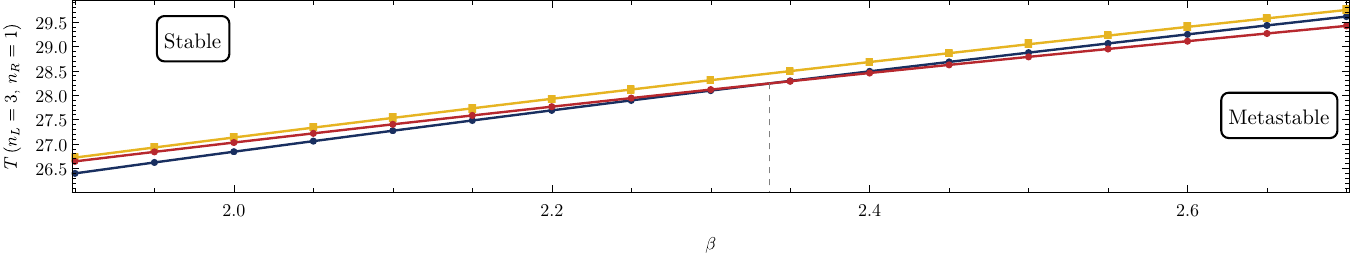}
\vskip 0.2cm
\includegraphics[width=\columnwidth]{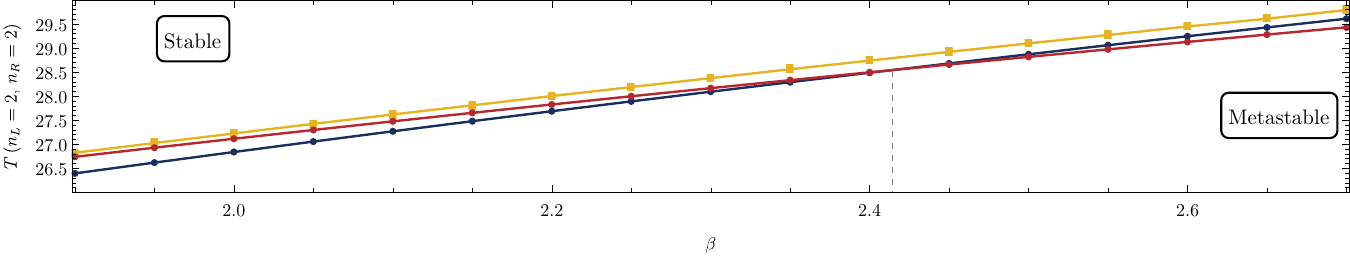}
\vskip 0.2cm
\includegraphics[width=\columnwidth]{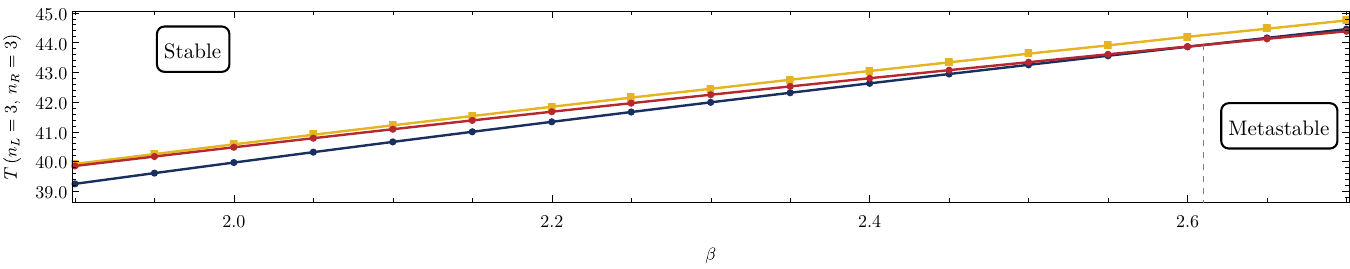}
\caption{\small
Stability and metastability of the CW-ANO string for different winding numbers $(n_L, n_R)$.
The blue and red lines show the total energy of the system at $d = 0$ and $d = \infty$, respectively, while the yellow line is the value of the energy barrier.
In the parameter region in which the red line is above (below) the blue line, the string is stable (metastable).
The winding numbers are $(n_L, n_R) = (1, 1)$, $(2, 1)$, $(3, 1)$, $(2, 2)$, and $(3, 3)$ from top to bottom.
}
\label{fig:stable-metastable}
\end{center}
\end{figure}

We examine whether the same behavior as in the previous subsection can be observed for strings with different winding numbers.
For this purpose we change the winding number of the left and right strings $n_L$ and $n_R$.
One example of the field configuration is shown in Fig.~\ref{fig:configuration_CW_largen} for $(n_L, n_R) = (3, 1)$.
Since the total winding number is $4$, the $\Phi$ field rotates four times along a circle at $r = \infty$.
Due to the larger winding number, the left string is thicker and develops a ring in the energy density~\cite{Vilenkin:2000jqa}.

Similarly to Sec.~\ref{subsec:numerical}, we solve the diffusion equation for different values of $\beta$ and $d$ to construct the interaction potential.
Fig.~\ref{fig:largen} shows the distance dependence of the total energy of the two-string system for winding numbers $(n_L, n_R) = (2, 1)$, $(2, 2)$, $(3, 1)$, and $(3,3)$.
We see that all the lines are monotonic for the AH case, while the barriers still remain for the CW case.

Finally, we show the stability-metastability diagram of the CW-ANO string for different winding numbers in Fig.~\ref{fig:stable-metastable}.
The blue and red lines are the string tension at $d = 0$ and $d = \infty$, respectively, while the yellow lines show the height of the energy barrier.
For $\beta$ smaller (larger) than the gray line, the tension at $d = 0$ is smaller (larger) than that at $d = \infty$, and thus the string with winding number $n = n_L + n_R$ can decay into two strings with $n_L$ and $n_R$.\footnote{
Note that the string decay can have multiple channels for $n \geq 4$.
For example, $n = 4 \to (n_L, n_R) = (3, 1)$ and $n = 4 \to (n_L, n_R) = (2, 2)$ are both allowed for $\beta \gtrsim 2.4$.
Also note that the decay chain can continue for more than one step, for example $4 \to (3, 1) \to (2, 1, 1) \to (1, 1, 1, 1)$.
The detailed construction of the decay channel is beyond the scope of this paper.
}

\subsection{Closer look at the $d$-dependence}
\label{subsec:closer}

\paragraph{AH case}
Let us take a closer look at the energy behavior we found in Sec.~\ref{subsec:numerical}.
Firstly, we consider the AH-ANO string. 
Let $T(d)$ be the minimized total tension for the two-string system with $(n_L,n_R)=(1,1)$ and $d$ being the fixed distance. 
The strings feel the repulsive (attractive) interaction potential when $T'(d)<0$ $(T'(d)>0)$.
It is convenient to introduce the tension difference
\begin{align}
\Delta T (d)
&\equiv
T (d) - T(d = 0).
\end{align}
As discussed in Sec.~\ref{sec:axisymmetric}, the AH-ANO string exhibits a special property in the BPS limit $\beta=1$ that $\Delta T(d)$ is independent of $d$, as each string has the translational moduli parameter.
For $\beta\neq 1$, it is difficult to investigate $\Delta T(d)$ analytically.
Thus we focus on two extreme cases: large $d$ ($\gg 1$) and small $d$ ($\ll 1$), and investigate its asymptotic behaviors.
The qualitative behavior with large $d$, i.e., for well-separated strings, is relatively easy to understand.
As studied in Sec.~\ref{sec:axisymmetric}, the string configurations take the asymptotic behaviors \eqref{eq: asymptotic} at large distances from the strings, and hence the interaction is dominated by the overlap between the exponential tails from the gauge field (scalar field) for $\beta>1$ ($\beta<1$).
Furthermore, the gauge and scalar fields provide the repulsive and attractive interactions, respectively~\cite{Vilenkin:2000jqa}.
Thus $\Delta T(d)$ at large $d$ increases and decreases in $d$ when $\beta<1$ and $\beta>1$, respectively.

On the other hand, the behavior of $\Delta T(d)$ for small $d$ is more complicated.
Thus we consider the near-BPS case, $\beta \simeq 1$, and perform the perturbative analysis with respect to $|\beta-1| \ll1$.
At the zero-th order of $\beta-1$, i.e., for the BPS limit $\beta =1$, the tension of the BPS solution reduces to the well-known result $4 \pi$ independent of $d$.
Furthermore, the BPS solution with $d=0$, i.e., the axisymmetric BPS solution with $n=2$, is given by
\begin{equation}
\Phi=f_\mathrm{BPS}(r) e^{2i\theta},
\label{eq: BPS sol d=0}
\end{equation}
with $f(r)_\mathrm{BPS} \propto r^2$ for $r \simeq 0$ and $f_\mathrm{BPS}(r) \to 1$ for $r \to \infty$.
Let us perturb the BPS solution \eqref{eq: BPS sol d=0} by an infinitesimal perturbation 
\begin{equation}
\delta_c \Phi = c h_2(r) f_\mathrm{BPS}(r),
\label{eq: perturbation} 
\end{equation}
with $c$ ($\ll 1$) an arbitrary real constant and $h_2$ satisfying the linearized EOM 
\begin{equation}
-\frac{1}{r}\frac{\df}{\df r}\left(r \frac{\df h_2}{\df r}\right) + \left( f_\mathrm{BPS}^2 + \frac{4}{r^2}\right) h_2 =0,
\label{eq: eom for h_2}
\end{equation}
from which one can see that $h_2(r)$ is approximately expanded $h_2(r) \propto r^{-2}$ for $r\simeq 0$.
This perturbation corresponds to splitting the axisymmetric solution into the two strings.
This can be seen by taking $c$ such that
\begin{equation}
c h_2 (r) \simeq - \frac{d^2}{4 r^2} \quad (\text{for } r\simeq 0),
\end{equation}
leading to the perturbed configuration
\begin{align}
\Phi + \delta_c \Phi &= f_\mathrm{BPS}(r) e^{2 i \theta} + c h_2(r) f_\mathrm{BPS}(r) ,
\end{align}
whose absolute value has two zero's at $(x,y)=(\pm d/2 ,0)$ as
\begin{align}
|\Phi + \delta_c \Phi|^2 &= f_\mathrm{BPS}(r)^2  \left[1+ c^2 h_2 (r) ^2 + 2 c h_2(r) \cos 2\theta \right]\label{eq: squared deformed phi} \\
&\simeq r^4  \left[1 - \frac{1}{2}\frac{d^2}{r^2} \cos 2\theta +\left(\frac{d}{2r}\right)^4\right] \quad (\text{for $r \simeq 0$}).
\end{align}
Note that, this perturbation does not change the tension from $4\pi$ to the order of $\mathcal{O}(c^2)$~\cite{Vilenkin:2000jqa}, and thus, to the order of $\mathcal{O}(d^4)$, the perturbed configuration coincides with the BPS solution with the nonzero interstring distance $d$.
In other words, this perturbation is nothing but the moduli, with respect to which the tension changes only by the order of $\mathcal{O}(c^3)$ ($\mathcal{O}(d^6)$).

\begin{figure}
\begin{center}
\includegraphics[width=0.5\columnwidth]{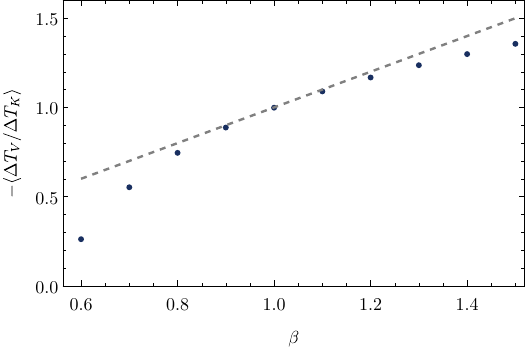}
\caption{\small
Ratio between $\Delta T_K$ and $\Delta T_V$ for the AH-ANO string.
The data points show the average value of $- \Delta T_V / \Delta T_K$ evaluated at $d = 0.1,0.2,\cdots, 1$.
They match well with the prediction from Eq.~\eqref{eq: relation T_K T_V}, i.e. $\langle \Delta T_V / \Delta T_K \rangle = - \beta + {\cal O} ((\beta - 1)^2)$.
The gray line is $\langle \Delta T_V / \Delta T_K \rangle = - \beta$ for comparison.
}
\label{fig:AH_ratio}
\end{center}
\end{figure}

\begin{figure}
\begin{center}
\includegraphics[width=\columnwidth]{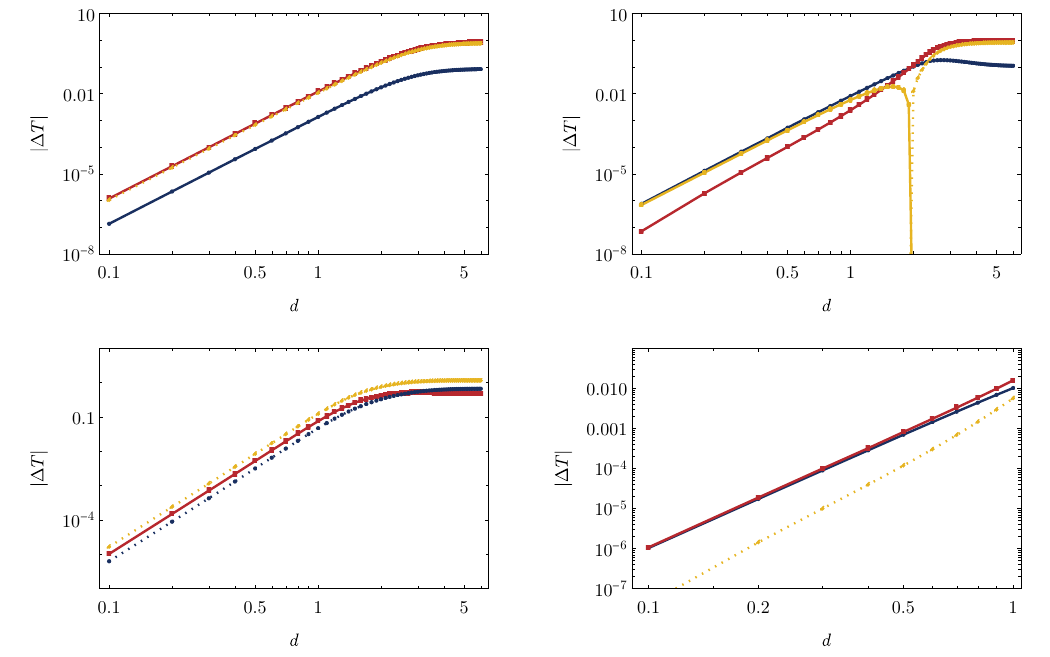}
\caption{\small
Behavior of the absolute value of the energy difference $\Delta T (d) = T (d) - T(d = 0)$ for the AH-ANO and CW-ANO strings.
The left panels show the AH case with $\beta = 0.9$ (top) and $\beta = 1.5$ (bottom), while the right panels show the CW case with $\beta = 1.8$ (top) and $\beta = 3$ (bottom).
The red and yellow lines are $|\Delta T_K (d)|$ and $|\Delta T_V (d)|$, respectively, while the blue lines are $|\Delta T (d)|$.
For the solid (dotted) lines, the quantity before taking the absolute value is positive (negative).
See also Figs.~\ref{fig:AH_detail} and \ref{fig:CW_detail}.
}
\label{fig:analytic_d4}
\end{center}
\end{figure}

Let us take into account the leading order of $\beta-1(\neq 0)$.
The tension is decomposed as
\begin{align}
T
&=
T_V + T_K,
\label{eq: tension}
\\[0.2cm]
T_V = \frac{\beta}{2} \int \df^2 x \, (|\Phi |^2-1)^2,
&\qquad  T_K = \int \df^2 x \, \left( \frac{1}{2}B_z^2 + |D_i \Phi|^2 \right),
\label{eq: T_K T_V}
\end{align}
where $T_V$ and $T_K$ are the contributions to the tension from the potential energy and from the sum of the kinetic energy for the scalar and gauge fields, respectively.
$T_K$ can be rewritten as
\begin{align}
T_K &= 4 \pi + \int \df^2 x \, \left[ \frac{1}{2} \left(B_z + |\Phi|^2-1\right)^2 + \left|(D_x + i D_y)\Phi\right|^2\right] - \frac{1}{2}\int \df^2 x \, (|\Phi |^2-1)^2,
\label{eq: T_K}
\end{align}
where we have used $\int \df^2 x B_z = 4 \pi$.
If the BPS limit is exact $\beta=1$, the third term in $T_K$ cancels with $T_V$, and thus the tension is minimized to be $4 \pi$ if and only if the second term in Eq.~\eqref{eq: T_K} vanishes, leading to the well-known BPS equations as studied in Sec.~\ref{sec:axisymmetric}.
Since $\beta$ now slightly differs from unity, the cancellation is not exact, and the solution for each fixed $d$ deviates from the BPS solutions by the order of $\mathcal{O}(\beta-1)$, say, $\Phi = \Phi^\mathrm{(BPS)}+ \delta_\beta \Phi$ and $A_i = A_i^\mathrm{(BPS)}+ \delta_\beta A_i$.
(``$\delta_\beta$'' indicates deviations of the order of $\mathcal{O}(\beta-1)$.)
By substituting these and using that $\Phi^\mathrm{(BPS)}$ and $ A_i^\mathrm{(BPS)}$ solve the BPS equations, it is found that the second term in Eq.~\eqref{eq: T_K} gives $\mathcal{O}((\beta-1)^2)$ terms, 
\begin{align}
T_K &= 4 \pi + \mathcal{O}((\beta-1)^2) - \frac{1}{2}\int \df^2 x \, (|\Phi |^2-1)^2.
\label{eq: rewritten T_K}
\end{align}
Therefore, we obtain a relation for the kinetic energy and the potential energy,
\begin{equation}
\frac{\Delta T_V(d)}{\Delta T _K(d)} = -\beta + \mathcal{O}((\beta-1)^2)
\label{eq: relation T_K T_V}
\end{equation}
with $\Delta T_V(d) \equiv T_V (d) - T_V(0)$ and $\Delta T_K(d) \equiv T_K (d) - T_K(0)$, which holds for every solution with arbitrary $d$.
This relation can be confirmed by the numerical results, see Fig.~\ref{fig:AH_ratio}.
As seen from the left panels of Fig.~\ref{fig:analytic_d4}, this relation is realized by $\Delta T_K(d)$ being positive and $\Delta T_V(d)$ being negative.
Since $|\Delta T_V / \Delta T_K | >1$ ($<1$) for $\beta>1$ ($\beta<1$) from Eq.~\eqref{eq: relation T_K T_V}, the sign of their sum $\Delta T(d)$ changes at $\beta=1$.

Let us see how $\Delta T(d)$ depends on $d$ to the leading order of $\beta-1$.
To this end, it is convenient to rewrite $T$ using Eqs.~\eqref{eq: T_K T_V} and \eqref{eq: rewritten T_K} as 
\begin{align}
T = 4\pi + \frac{\beta-1}{2} \int \df^2 x \, (|\Phi |^2-1)^2 + \mathcal{O}((\beta-1)^2),
\end{align}
from which it can be seen that the deviations from the BPS solution, $\delta_\beta \Phi$ and $\delta_\beta A_i$, give negligible contributions of the order of $\mathcal{O}((\beta-1)^2)$ due to the factor $\beta-1$ in the second term.
Thus, to the leading order of $\beta-1$, we can take the configuration as the BPS solution $\Phi \simeq \Phi^\mathrm{(BPS)}$ and $A_i \simeq A_i^\mathrm{(BPS)}$.
In particular, the solution with $d=0$ can be taken as Eq.~\eqref{eq: BPS sol d=0}.

Then, we again perturb the solution by acting the infinitesimal deformation \eqref{eq: perturbation}.
The perturbed one coincides with the BPS solution with fixed $d$ to the order of $\mathcal{O}(d^4)$.
However, since $\beta$ deviates from unity, the perturbation is no longer the moduli of the tension.
Instead, the perturbation changes the tension as
\begin{align}
T &= \frac{\beta-1}{2} \int \df^2 x \, \left( f_\mathrm{BPS}(r)^2 \left[ 1 + c^2 h_2 (r) ^2 + 2 c h_2(r) \cos 2\theta \right] - 1 \right)^2 \nn
& \hspace{4em} + 4\pi + \mathcal{O}((\beta-1)^2),
\end{align}
and hence
\begin{align}
\Delta T(d) & = (\beta-1) c^2 \int \df^2 x \, h_2 (r)^2 \left( f_\mathrm{BPS}(r)^2-1 \right) + \mathcal{O}(c^4) + \mathcal{O}((\beta-1)^2),
\label{eq: delta T leading}
\end{align}
where we have used Eq.~\eqref{eq: squared deformed phi} and $\int \df\theta \cos 2\theta=0$.
Thus the tension changes by the order of $\mathcal{O}(c^2)$ ($\mathcal{O}(d^4)$), instead of $\mathcal{O}(c^3)$ ($\mathcal{O}(d^6)$).
From Eq.~\eqref{eq: delta T leading}, it can be seen that the sign of $\Delta T(d)$ changes from positive to negative as $\beta$ exceeds unity (note that $f_\mathrm{BPS}(r)^2 <1$ holds everywhere), which agrees with the behavior derived from the relation \eqref{eq: relation T_K T_V}.

From the above analysis for small $d$,
it follows that $\Delta T(d)$ increases and decreases as $\propto \pm d^4$ for $\beta< 1$ and $\beta> 1$, respectively.
Note that this analytical argument is valid to the leading order of $\mathcal{O}(\beta-1)$.
Remarkably, this behavior is confirmed by the numerical calculation even for wide range of $\beta$, see the left panels in Fig.~\ref{fig:analytic_d4}.
The top-left and bottom-left panels show the behavior of $\Delta T(d)$ at small $d$ for $\beta=0.9$ and $\beta = 2.0$, respectively.
The former one has the asymptotic behavior increasing with $\propto d^4$ while the latter one has decreasing one with $\propto d^4$.

Therefore, the asymptotic behavior of $\Delta T(d)$ in the AH case is summarized as below:
\begin{itemize}
    \item $\Delta T(d)$ increases/decreases with exponential behaviors at large $d$ for $\beta<1$ $(\beta>1)$.
    \item $\Delta T(d)$ increases/decreases being proportional to $d^4$ at small $d$ for $\beta<1$ $(\beta>1)$.
\end{itemize}
These agree well with the qualitative structure read off from the left panels in Fig.~\ref{fig:d>0_beta-E}, which do not have any non-trivial energy barrier.


\paragraph{CW case}

In the case of the CW-ANO string, in contrast, there is no simple way of understanding the behavior of the string tension like the AH-ANO string.
We again focus on the two extreme cases: large $d$ and small $d$.
For large $d$, the asymptotic behavior of $\Delta T(d)$ is the same as that of the AH case, because the analysis that led to Eq.~\eqref{eq: asymptotic} holds independently of the detailed shape of the potential.
Thus $\Delta T(d)$ exponentially increases and decreases at large $d$ when $\beta < 1$ and $\beta > 1$, respectively.

On the other hand, the small $d$ behavior is much more complicated.
In the blue lines of Fig.~\ref{fig:analytic_d4} we show the absolute value of the tension difference $|\Delta T (d)|$ at small $d$.
We also show its kinetic and potential contributions $|\Delta T_K (d)|$ and $|\Delta T_V (d)|$ in red and yellow, respectively.
In this plot, solid (dotted) lines mean that the quantity before taking the absolute value is positive (negative).
From the top-right ($\beta = 1.8$) and bottom-right ($\beta = 3$) panels, we see that the CW-ANO string behaves as $|\Delta T (d)| \propto d^4$ with positive coefficients around $d \sim 0$ independently of the value of $\beta$.
However, their kinetic and potential fractions are totally different.
For $\beta = 1.8$, it is the potential contribution $\Delta T_V (d)$ that dominates the $d$-dependence for small $d$, while it is the kinetic $\Delta T_K (d)$ for $\beta = 3$.
Also, the sign of $\Delta T_V (d)$ changes between these two panels.
In Fig.~\ref{fig:CW_detail} in App.~\ref{app:CW_detail} we show how $\Delta T_K (d)$ and $\Delta T_V (d)$ behave for different values of $\beta$.
We clearly see the tendency that $\Delta T_V (d)$ dominates the behavior of $\Delta T (d)$ for small $\beta$, while $\Delta T_K (d)$ starts to dominate as $\beta$ increases.

The energy barrier for the CW-ANO string appears as a result of these asymptotic behaviors.
The existence of the barrier requires $\Delta T (d)$ be an increasing and deceasing functions of $d$ for small and large distances, respectively.
The latter is guaranteed for $\beta > 1$, while the former is difficult to understand analytically.

\section{Discussion and conclusions}
\label{sec:dc}

In this paper we have investigated the properties of the Abelian-Higgs string described by the CW potential.
As well known, the Abelian-Higgs string described by the usual quadratic-quartic potential (which we simply call the AH-ANO string) has two phases, type I (attractive) and type II (repulsive), and the only deterministic parameter is the mass ratio of the gauge boson to the Higgs $\beta = m_\Phi^2 / m_A^2$, with the BPS state existing at the boundary $\beta = 1$.
However, in high-energy physics, this is not the only potential that leads to spontaneous symmetry breaking in low-energy effective theories.
One typical example is the Coleman-Weinberg (CW) potential.
While the $\lambda \Phi^4$ potential classically admits only the trivial vacuum with no string, nontrivial vacua arise once quantum corrections are taken into account.
We call the string realized by this potential the CW-ANO string, and have investigated in detail the difference of its properties from those of the usual AH-ANO string.

In Sec.~\ref{sec:axisymmetric} we have estimated the energies for different mass ratios $\beta$ and winding numbers $n$ for axisymmetric strings.
For the AH-ANO string, the energy per unit length (tension) for different $n$ intersects at a single point (BPS) at $\beta = 1$ (see Fig.~\ref{fig:d_0_beta-E}), whereas this is not the case for the CW-ANO string.
This fact already suggests that the phase diagram of the CW-ANO string has a richer structure.
We have also investigated the field configurations for both AH-ANO and CW-ANO strings.
Although the fractions of the kinetic and potential contributions are not significantly different between the two, some differences have been observed: For the CW-ANO string, the values of both contributions themselves are smaller, and radius at which the configuration mainly contributes to the total energy is located outward.
These reflect the flatness of the potential near the origin.

In order to investigate the richer structure of the CW-ANO string suggested by the analysis of axisymmetric strings, in Sec.~\ref{sec:two_string} we have numerically determined the interaction potential (minimum energy) of the two-string system as a function of the interstring distance $d$.
This can be done without numerical difficulty by rewriting the equation for the minimum energy as a diffusion equation (Eq.~(\ref{eq: diff eq 2d}), also called the flow equation).
Interestingly, an energy barrier has been observed in the interaction potential at some distance for CW-ANO strings with $\beta > 1$ (Fig.~\ref{fig:d>0_beta-E}), meaning that the interstring force is attractive (repulsive) below (above) that distance.
While such coexistence of attraction and repulsion is known for so-called type-1.5 strings, the attractive/repulsive relation is found to be opposite in the CW-ANO string.
Thus we call the latter type-$\overline{1.5}$.
We have also found that the relative magnitude of the energy at $d = 0$ and at $d = \infty$ depends on the value of $\beta$, and that the transition occurs at some critical value $\beta = \beta_c$, not at $\beta = 1$.
Such transition in string properties at multiple values of $\beta$ is one distinct feature not observed in AH-ANO strings, and it suggests that vortex strings in general have richer properties than previously thought.

We also have had a closer look at the energy barrier observed in the CW-ANO string in Sec.~\ref{subsec:closer}.
For small $d$, the kinetic and potential terms contribute to the total energy of the AH-ANO string as increasing and decreasing functions of $d$, respectively.
Their relative magnitude changes at $\beta = 1$, so that the attractive/repulsive relation also changes across this value.
In the case of the CW-ANO strings, however, the $d$-dependence of the string tension is much more complicated, due to the absence of the BPS state.
While the interstring force at large $d$ behaves in a similar way as the AH-ANO string (i.e. attractive (repulsive) for $\beta < 1$ ($\beta > 1$), see also the last paragraph of Sec.~\ref{sec:ANO string solution}), it behaves nontrivially at small $d$ (Figs.~\ref{fig:analytic_d4} and \ref{fig:CW_detail}). The combination of different behavior at small and large $d$ results in the appearance of the energy barrier in the CW-ANO string.

In App.~\ref{app:universality}, we have confirmed the same features for other potentials that have a flat structure around the origin.
As mentioned in Sec.~\ref{subsec:validity}, the analysis of the CW-ANO string in this paper is not fully justified in that it uses the effective potential alone, which is merely the leading term of the effective action.
However, the universality confirmed at least suggests that strings with different potentials have rich phase structures than previously thought.

In our analysis, we did not specify the origin of the logarithmic running of the quartic coupling $\lambda(\Phi)$, which could be radiatively generated by loop effects of scalar bosons, gauge bosons, or fermions in general.
There still remains the question whether the same result is obtained without integrating out these underlying particles.
This is an interesting but highly non-trivial question, which will be tackled elsewhere.

There are many possible applications of the analysis in this paper.
Relatively straightforward applications would be to investigate string properties for a larger variety of potentials, or to compute the quantum decay of strings.
Other applications include going beyond the simplest Abelian-Higgs model.
When the $U(1)$ gauge field is coupled to more complex scalar fields (the extended Abelian-Higgs model), strings are called semi-local strings \cite{Vachaspati:1991dz,Achucarro:1999it}.
The extension of our work to the case of semi-local strings is one of interesting future directions.
On the other hand, when the $U(N)$ gauge field is coupled to $N \times N$ matrix complex scalar field (the non-Abelian Higgs model), strings are called non-Abelian strings \cite{Hanany:2003hp,Auzzi:2003fs,Eto:2005yh,Eto:2006cx} (for $N \times N_f$ ($N_f >N$) matrix scalar field, then non-Abelian semi-local strings \cite{Shifman:2006kd,Eto:2007yv}).
Non-Abelian strings have internal orientational moduli and have been studied extensively, see Refs.~\cite{Tong:2005un,Eto:2006pg,Shifman:2007ce,Shifman:2009zz,Tong:2008qd} for a review.
Apparently, a non-Abelian extension of our work is also worth studying.

As for applications to high-energy phenomenology, it is worth to point out that electroweak $Z$ strings are discussed in the Standard Model (SM) \cite{Nambu:1977ag,Vachaspati:1992fi,Achucarro:1999it}.
These strings are nontopological and in fact are unstable in the realistic parameter region \cite{James:1992zp,James:1992wb,Goodband:1995he,Achucarro:1999it}.
The same happens for models beyond the SM (BSM) such as two-Higgs doublet models \cite{Earnshaw:1993yu,Eto:2021dca} (see also Refs.~\cite{Dvali:1993sg,Eto:2018hhg,Eto:2018tnk,Eto:2019hhf,Eto:2020hjb,Eto:2020opf} for topological fractional $Z$-strings).
It is an interesting open question whether a CW-type potential can stabilize $Z$-strings in the (B)SM.
From the viewpoint of cosmology, cosmic strings are one of the interesting targets of the ongoing and future gravitational wave observatories~\cite{NANOGrav:2020qll,Desvignes:2016yex,Kerr:2020qdo,Yagi:2011wg,LISA:2017pwj,Taiji,TianQin:2020hid,Punturo:2010zz,Sesana:2019vho,AEDGE:2019nxb}, and the nontrivial dependence of the energy on the string distance may affect the reconnection dynamics and thus leave its own characteristic imprint on the spectrum.
We leave such a study for future work.

We conclude this section by mentioning another interesting implication for condensed matter physics.
If one can realize a superconducting material that is described by an effective theory similar to the one we studied in this paper, i.e., the Landau-Ginzburg theory with the CW potential without the quadratic term, one would observe a non-trivial behavior of the vortices for $\beta >1$.
For example, consider an external magnetic field applied to the material.
When the magnetic field is relatively weak but suffices to penetrate it, the vortices are dilute and the typical distance between neighboring vortices is large.
At this point the interaction is repulsive, just in the same way as the ordinary Abrikosov lattice.
However, as the magnetic field gets stronger, the number of the vortices increases and the typical distance between them gets smaller.
As a result, once the magnetic field exceeds a critical value, the distance between some of the neighboring vortices becomes so small that the interaction between them flips the sign and the pairs start to merge into vortices with winding number two.
What would happen if we make the magnetic field stronger?
One possibility is that the material behaves similarly as buffer solution: after some of the pairs merge, the distance between vortices would be large enough again for the vortices to feel the repulsive force.
As the magnetic field becomes further stronger, the number of the merged pairs increases, but the lattice structure would still remain.
Therefore, such a material can be more stable against the magnetic field than the conventional type-II superconductors.

\section*{Acknowledgements}
The authors would like to thank Kohei Fujikura for useful comments.
The work of M.\,E. and M.\,N. is supported in part by JSPS Grant-in-Aid for
Scientific Research (KAKENHI Grant No.~JP22H01221).
The work of M.\ E. is supported in part by the JSPS Grant-in-Aid for Scientific Research 
KAKENHI Grant No.~JP19K03839
and the MEXT KAKENHI Grant-in-Aid for Scientific Research on Innovative Areas
``Discrete Geometric Analysis for Materials Design'' No.~JP17H06462 from the MEXT of Japan.
The work of Y.\ H. is supported in part by the JSPS Grant-in-Aid for Scientific Research 
KAKENHI Grant No.~JP21J01117.
The work of R.\,J. is supported by the grants IFT Centro de Excelencia Severo Ochoa SEV-2016-0597, CEX2020-001007-S and by PID2019-110058GB-C22 funded by MCIN/AEI/10.13039/501100011033 and by ERDF.
The work of R.\,J. is supported by the Deutsche Forschungsgemeinschaft under Germany's Excellence Strategy -- EXC 2121  ``Quantum Universe'' -- 390833306.
The work of R.\,J. is supported by Grants-in-Aid for JSPS Overseas Research Fellow (No.~201960698).
The work of M.\ N. ~is supported in part by JSPS Grant-in-Aid for
Scientific Research (KAKENHI Grant No.~JP18H01217).
The work of M.\,Y. is supported by the DFG Collaborative Research Centre ``SFB 1225 (ISOQUANT)'', Germany’s
Excellence Strategy EXC-2181/1-390900948 (the Heidelberg Excellence Cluster STRUCTURES) and the Alexander von Humboldt Foundation.

\appendix

\section{Universality of the barrier for flat potentials}
\label{app:universality}

\begin{figure}
\begin{center}
\includegraphics[width=0.6\columnwidth]{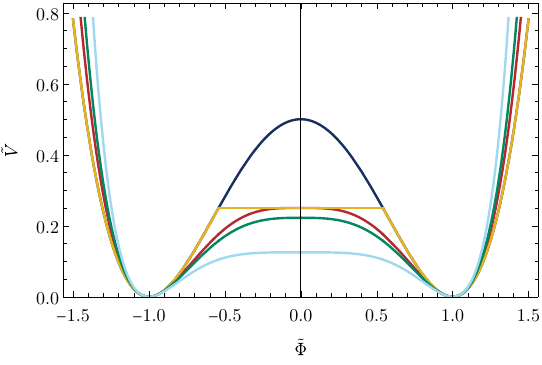}
\caption{\small
Potentials used in App.~\ref{app:universality}: $\tilde{V}_{\rm AH}$ (blue), $\tilde{V}_{\rm CW}$ (red), $\tilde{V}_{\rm AH-cut}$ with $\tilde{V}_0 = 0.5$ (yellow), $\tilde{V}_{{\rm AH}-36}$ (green), and $\tilde{V}_{{\rm AH}-48}$ (light blue).
}
\label{fig:V}
\end{center}
\end{figure}

\begin{figure}
\begin{center}
\includegraphics[width=\columnwidth]{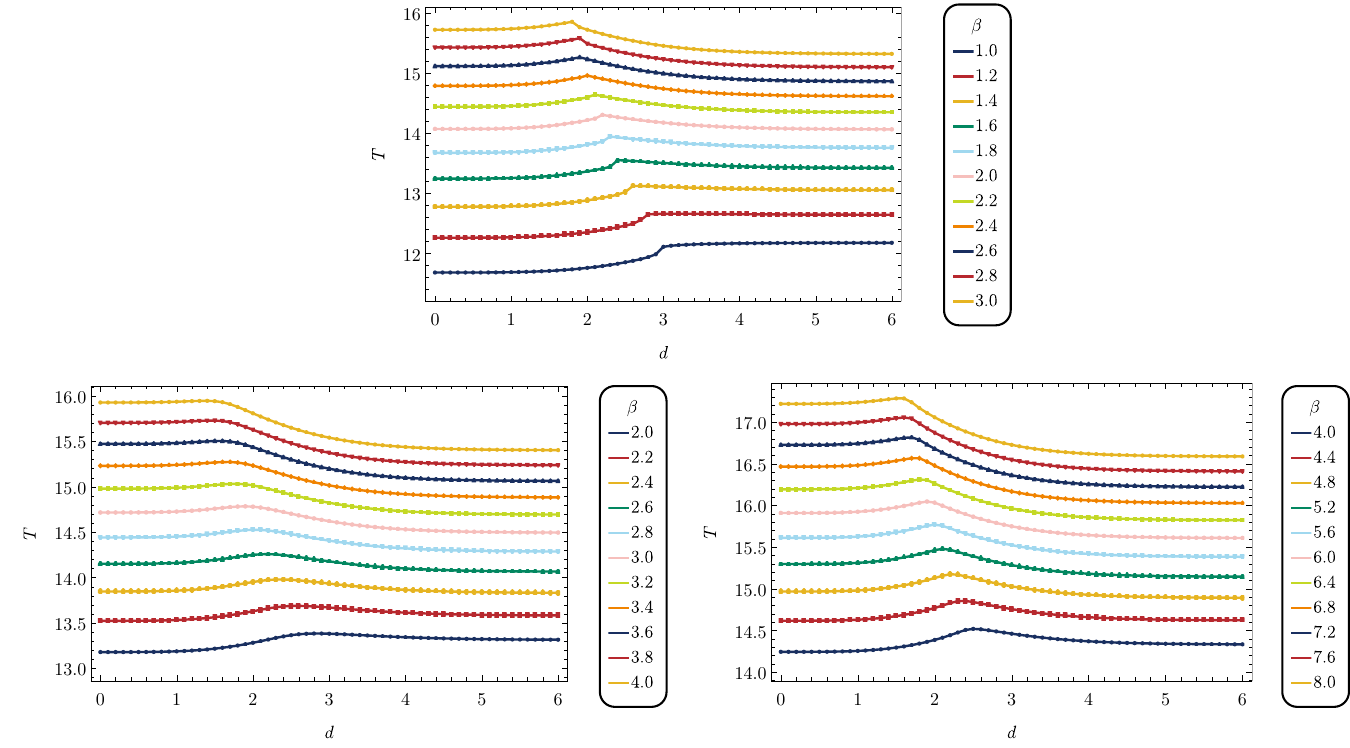}
\caption{\small
Distance dependence of the energy of the two-string system with the $\tilde{V}_{\rm AH-cut}$ with $\tilde{V}_0 = 0.5$ (top), $\tilde{V}_{\rm AH-36}$ (bottom-left), and $\tilde{V}_{\rm AH-48}$ (bottom-right). 
}
\label{fig:d>0_beta-E_others}
\end{center}
\end{figure}

In this appendix, we analyze several types of potentials that have flat structure around the origin, and show that the strings have similar properties as we found in the main text for the CW potential.
We study the rescaled potentials
\begin{align}
\tilde{V}_{\rm AH-cut}
&=
\left\{
\begin{array}{ll}
\displaystyle
\frac{\beta}{2} \tilde{V}_0
&\quad
\left( |\tilde{\Phi}| < \sqrt{1 - \sqrt{\tilde{V}_0}} \right),
\\[4ex]
\displaystyle
\frac{\beta}{2} \left( |\tilde{\Phi}|^2 - 1 \right)^2
&\quad
\left( |\tilde{\Phi}| > \sqrt{1 - \sqrt{\tilde{V}_0}} \right),
\end{array}
\right.
\\[2ex]
\tilde{V}_{\rm AH-36}
&=
\frac{2 \beta}{9} \left( |\tilde{\Phi}|^3 - 1 \right)^2,
\\[2ex]
\tilde{V}_{\rm AH-48}
&=
\frac{\beta}{8} \left( |\tilde{\Phi}|^4 - 1 \right)^2.
\end{align}
We plot these potentials in Fig.~\ref{fig:V}.
We calculate the $d$-dependence of the energy of the two-string system with the same method as the main text.
The result is shown in Fig.~\ref{fig:d>0_beta-E_others}.
We see that all of these potentials develop an energy barrier around $d \sim 2 - 3$.


\section{Explanation with one-string ansatz}
\label{app:ansatz}

\begin{figure}
\begin{center}
\includegraphics[width=\columnwidth]{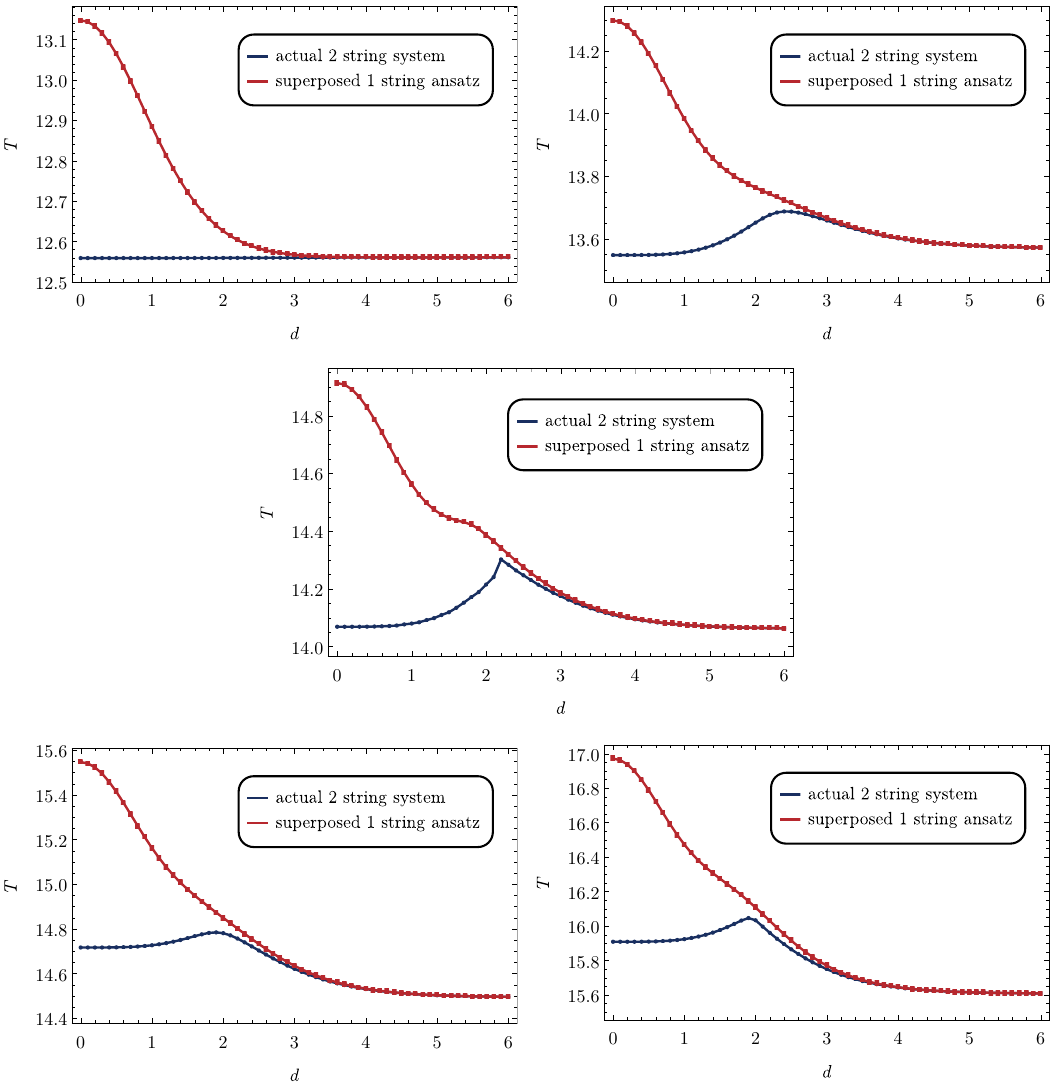}
\caption{\small
Comparison between the energy of the actual two-string system (blue) and superposed one-string ansatz (red).
The potential is $V_{\rm AH}$ with $\beta = 1$ (top-left), $V_{\rm CW}$ with $\beta = 2$ (top-right), $V_{\rm AH-cut}$ with $\beta = 2$ (middle), $V_{{\rm AH}-36}$ with $\beta = 3$ (bottom-left), and $V_{{\rm AH}-48}$ with $\beta = 6$ (bottom-right).
}
\label{fig:d>0_ansatz}
\end{center}
\end{figure}

In this appendix we study whether the superposition of the string configuration with winding number $n = 1$ can explain the behavior of the energy.
In Fig.~\ref{fig:d>0_ansatz} we show the comparison between the energy calculated from the actual two-string configuration (blue) and that from superposed one-string ansatz (red).
For the latter, we first solve for the axisymmetric configuration that minimizes the energy for $n = 1$, and then superpose two of such configurations at distance $d$.
The superposition is done with $\Phi = \Phi_L \times \Phi_R$ and $A_i = A_{i,L} + A_{i,R}$, with the subscripts $L$ and $R$ are for the left and right strings, respectively.
We see that the superposed configurations explain the behavior of the blue lines at large distances, while they fail at small distances.

\section{Details for the string tension}
\label{app:CW_detail}

\begin{figure}
\begin{center}
\includegraphics[width=0.8\columnwidth]{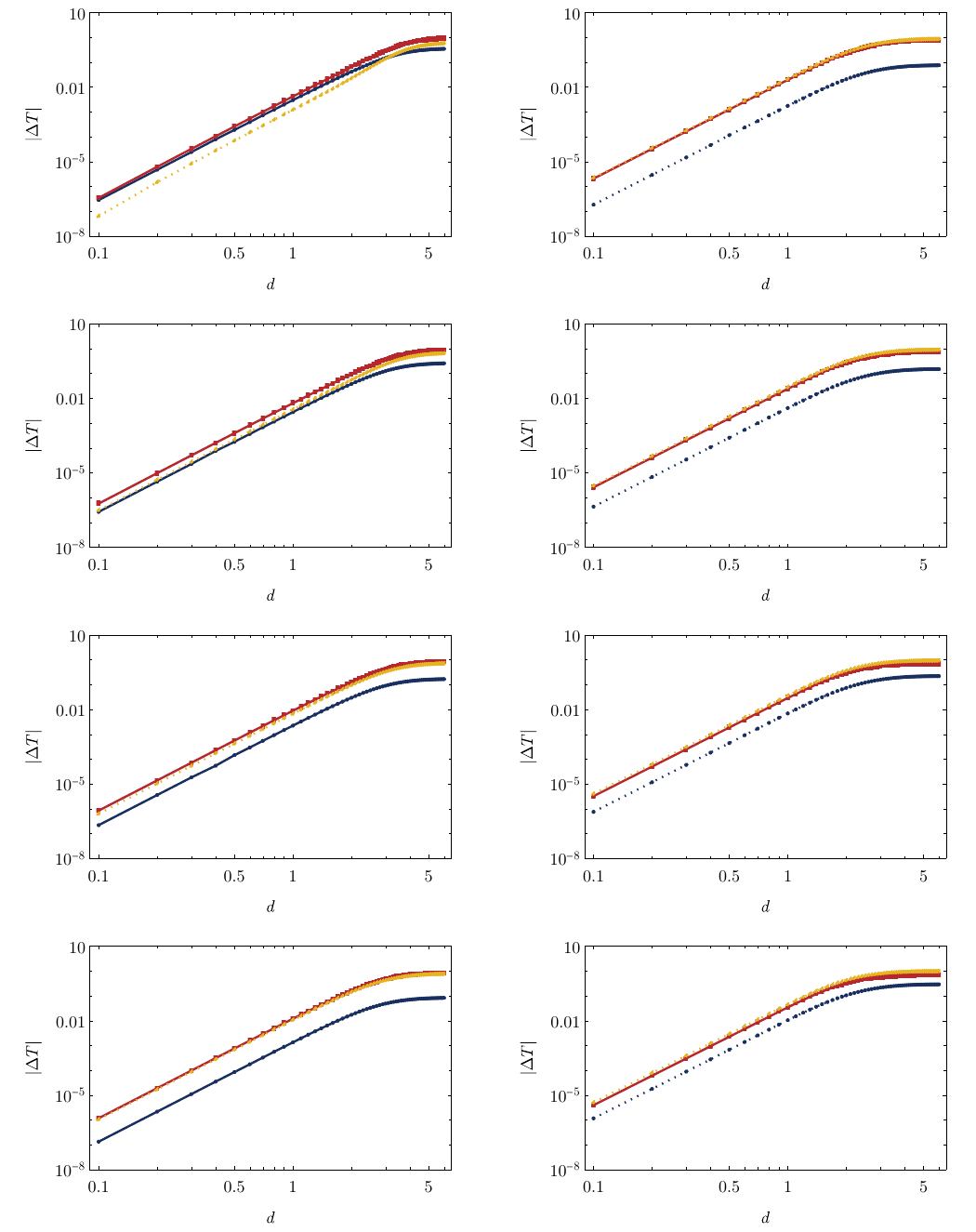}
\caption{\small
$|\Delta T_K|$ (red), $|\Delta T_V|$ (yellow), and $|\Delta T|$ (blue) for the AH-ANO string.
The value of $\beta$ is $\beta = 0.6, 0.7, 0.8, 0.9$
(left column) and $\beta = 1.1, 1.2, 1.3, 1.4$ (right column) from top to bottom.
The solid (dotted) lines mean that the quantity before taking the absolute value is positive (negative).
}
\label{fig:AH_detail}
\end{center}
\end{figure}

\begin{figure}
\begin{center}
\includegraphics[width=0.8\columnwidth]{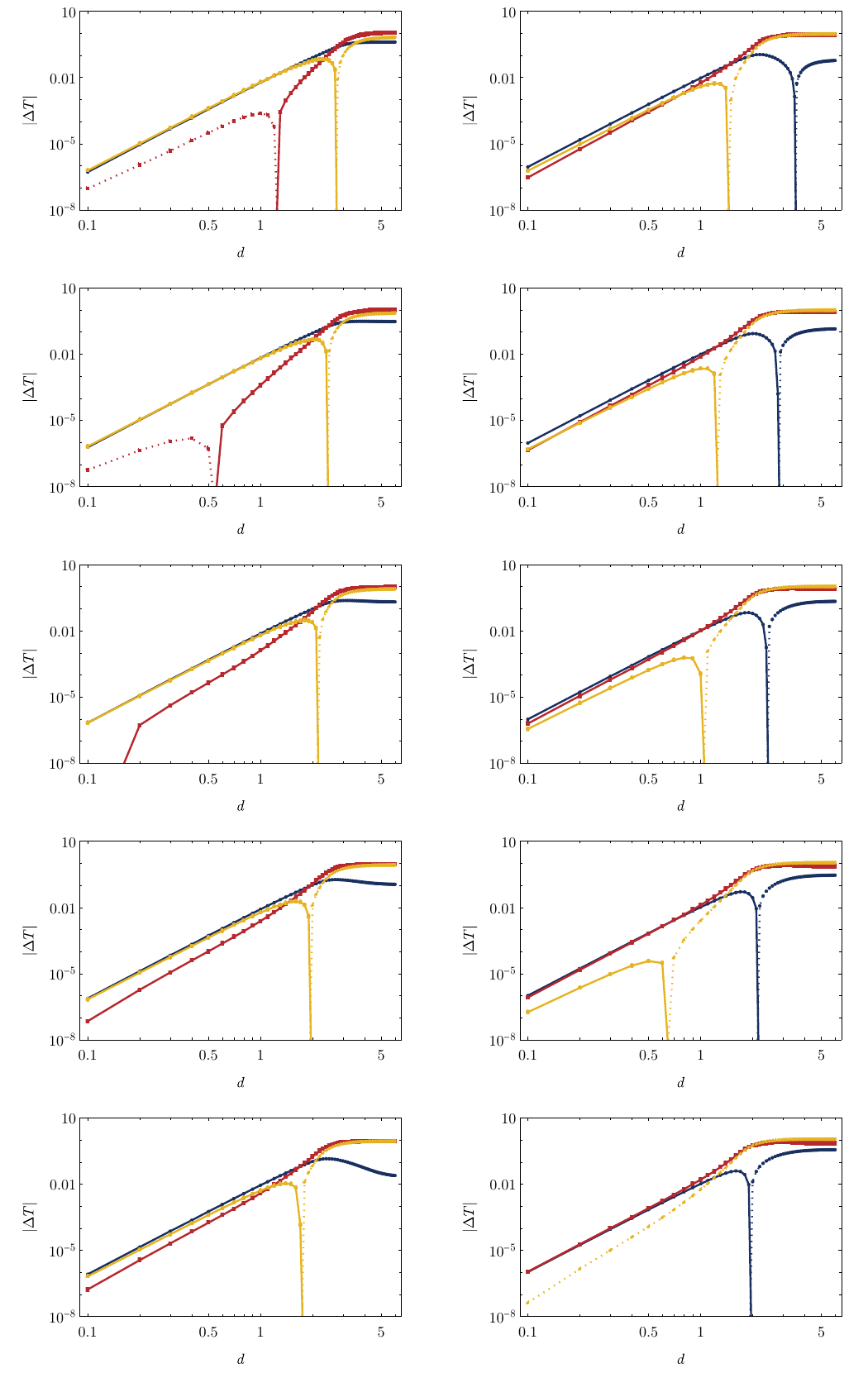}
\caption{\small
$|\Delta T_K|$ (red), $|\Delta T_V|$ (yellow), and $|\Delta T|$ (blue) for the CW-ANO string.
The value of $\beta$ is $\beta = 1.2, 1.4, 1.6, 1.8, 2.0$ (left column) and $\beta = 2.2, 2.4, 2.6, 2.8, 3.0$ (right column) from top to bottom.
The solid (dotted) lines mean that the quantity before taking the absolute value is positive (negative).
}
\label{fig:CW_detail}
\end{center}
\end{figure}

In this appendix we show the kinetic and potential contributions $\Delta T_K(d)$ and $\Delta T_V(d)$ to the string tension difference $\Delta T(d)$ for a wider range of parameter values.
Figs.~\ref{fig:AH_detail} and \ref{fig:CW_detail} show how the string tension $\Delta T_K(d)$, $\Delta T_V(d)$, and $\Delta T(d)$ behave for different values of $\beta$ and $d$ for the AH-ANO and CW-ANO strings, respectively.
The blue lines are the absolute value of the tension difference $|\Delta T (d)|$, while the red and yellow lines are its kinetic and potential contributions $|\Delta T_K (d)|$ and $|\Delta T_V (d)|$.
The solid and dotted lines indicate that the quantity before taking the absolute value is positive and negative, respectively.

For the AH-ANO string, $\Delta T_K(d)$ and $\Delta T_V(d)$ are always positive and negative, respectively.
Their ratio is $\Delta T_V(d) / \Delta T_K(d) \simeq - \beta$ at the leading order in $\beta - 1$ (see Sec.~\ref{subsec:closer}).
Since their relative magnitude changes across $\beta = 1$, their sum $\Delta T(d)$ behaves as an increasing and decreasing function for $\beta < 1$ and $\beta > 1$, respectively.

In contrast, $\Delta T_K(d)$ and $\Delta T_V(d)$ for the CW-ANO string can change the sign at some value of $d \lesssim {\cal O} (1)$, depending on the value of $\beta$.
While the asymptotic behavior for $d \gg {\cal O} (1)$ can be understood analytically (see Sec.~\ref{subsec:closer}), the behavior at small $d$ is hard to grasp.
The energy barrier for the CW-ANO string appears as a result of these asymptotic behaviors.


\clearpage

\bibliographystyle{JHEP} 
\bibliography{refs}
\end{document}